\documentclass[prd,nofootinbib,preprint,superscriptaddress]{revtex4}
\pdfoutput=1
\usepackage[T1]{fontenc}
\usepackage{amsmath,amssymb}
\usepackage{braket}
\usepackage{epsfig}
\usepackage{graphicx}
\usepackage[usenames,dvipsnames]{color}
\usepackage{subfigure}
\usepackage{slashed}
\usepackage[colorlinks,citecolor=blue]{hyperref}
\usepackage{pdfpages}
\usepackage{color}
\usepackage{comment}
\begin{document}
%\preprint{IP/BBSR/2015-4}
%\title{Imprints of ultralight PBH formed during phase transitions on gravitational waves}

%\title{Probing ultra-light PBH formation mechanism via unique doubly-peaked gravitational waves spectrum}

\title{Axion misalignment with memory-burdened PBH}

\author{Disha Bandyopadhyay}
\email{b.disha@iitg.ac.in}
\affiliation{Department of Physics, Indian Institute of Technology Guwahati, Assam 781039, India}

\author{Debasish Borah}
\email{dborah@iitg.ac.in}
\affiliation{Department of Physics, Indian Institute of Technology Guwahati, Assam 781039, India}
\affiliation{Pittsburgh Particle Physics, Astrophysics, and Cosmology Center, Department of Physics and Astronomy, University of Pittsburgh, Pittsburgh, PA 15260, USA}

\author{Nayan Das}
\email{nayan.das@iitg.ac.in}
\affiliation{Department of Physics, Indian Institute of Technology Guwahati, Assam 781039, India}

\begin{abstract}
We study the possibility of producing axion dark matter (DM) via misalignment mechanisms in a non-standard cosmological era dominated by ultra-light primordial black holes (PBH). While the effect of PBH domination on the production of axion via vacuum misalignment is known assuming the PBH evaporation to proceed according to Hawking's semi-classical (SC) approximation, we go beyond these simplest possibilities to include kinetic misalignment of axion and backreaction effect of emitted particles on the PBH themselves, referred to as the memory-burden (MB) effect. We show that, depending upon the type of misalignment mechanism and PBH evaporation regime, the axion as well as PBH parameter space consistent with the observed DM relic changes significantly having interesting implications for axion detection experiments. PBH also offer complementary detection prospects via gravitational wave due to PBH density fluctuations and excess radiation due to emission of hot axions within reach of future cosmic microwave background experiments.
\end{abstract}
%\pacs{}
\maketitle
%%%%%%%%%%%%%%%%%%%%%%%%%%%%%%%%%%%%%%%%%%%%%%%%%%%%%%%%%%%%%%%%%
\section{Introduction}
\label{sec:Intro}
Axion \cite{Wilczek:1977pj, Weinberg:1977ma} is the pseudo Nambu-Goldstone boson (pNGB) associated with the Peccei-Quinn (PQ) symmetry $U(1)_{\rm PQ}$ \cite{Peccei:1977hh, Peccei:1977ur} introduced to solve the strong CP problem, a longstanding puzzle in particle physics. While electroweak interactions in the standard model (SM) has CP violations \cite{ParticleDataGroup:2020ssz}, the strong interaction counterpart characterized by the angle $\theta$ is tightly constrained $\theta<10^{-10}$ due to limits on the electric dipole moment of neutron \cite{Abel:2020pzs}. In the Peccei-Quinn mechanism \cite{Peccei:1977hh, Peccei:1977ur, Wilczek:1977pj, Weinberg:1977ma}, the pseudoscalar axion field dynamically determines $\theta$ keeping it naturally small. In other words, once the axion potential is generated via QCD effects, $\theta$ relaxes to zero, the ground state of the potential. In addition to providing an elegant solution to the strong CP problem, axion can also solve other longstanding puzzles in particle physics like the origin of dark matter (DM) and baryon asymmetry of the Universe (BAU) \cite{Planck:2018vyg, ParticleDataGroup:2020ssz}. While ultra-light axion with sub-eV mass can constitute the entire cold dark matter in the Universe \cite{Preskill:1982cy, Abbott:1982af, Dine:1982ah}, there are several ways in which axion or a $U(1)_{\rm PQ}$ framework can assist in generating the observed BAU \cite{Servant:2014bla, Ipek:2018lhm, Croon:2019ugf, Co:2019wyp}. Axions can also have interesting detection prospects and hence their couplings to the SM particles, decided by the PQ symmetry breaking scale $f_a$, remain tightly constrained. For QCD axion models with sub-eV axion mass, there exists a stringent astrophysical bound $f_a\gtrsim 10^{8\div9}$ GeV \cite{Raffelt:2006cw, Caputo:2024oqc}. Coherent oscillations of the axion field in an expanding Universe can lead the production of cold dark matter (CDM) within the framework of the vacuum misalignment mechanism \cite{Preskill:1982cy, Abbott:1982af, Dine:1982ah}. In order to produce the entire CDM observed in the present Universe, the decay constant is constrained to be around $f_a\sim 10^{11}$ GeV \cite{Kawasaki:2013ae}. Depending on the details of the QCD axion model namely, Dine-Fischler-Srednicki-Zhitnitsky (DFSZ) \cite{Dine:1981rt, Zhitnitsky:1980tq} or Kim-Shifman-Vainshtein-Zakharov (KSVZ) \cite{Kim:1979if, Shifman:1979if}, additional contributions from the topological defects can alter this window by less than an order of magnitude \cite{Kawasaki:2013ae}. If the axion field has an initial non-zero velocity, as considered in the kinetic misalignment mechanism \cite{Co:2019jts, Chang:2019tvx}, the decay constant $f_a$ can be much lower while being consistent with the observed CDM relic.

Since the details of the misalignment mechanism: vacuum or kinetic, decides the axion decay constant $f_a$, it can have interesting implications for axion detection prospects as axion-SM coupling is dictated by $f_a$. It is noteworthy that such bound on $f_a$ from CDM relic criteria is based on the assumption that the Universe was radiation dominated prior to the big bang nucleosynthesis (BBN) epoch. Presence of non-standard cosmological epochs in the early Universe can, therefore, change the bounds on $f_a$ from CDM relic criteria. We are interested in an early matter domination (EMD) phase\footnote{See \cite{Arias:2021rer, Visinelli:2017imh} for axion production via vacuum misalignment in a non-standard cosmological epoch with general equation of state.} which can arise either due to long-lived field \cite{Visinelli:2009kt,Nelson:2018via} or by ultra-light primordial black holes (PBH) \cite{Bernal:2021yyb, Mazde:2022sdx}. While both of these possible EMD phases can affect axion parameter space from the relic criteria in similar ways, we consider a PBH dominated EMD phase in the early Universe as it has other detectable signatures at gravitational wave (GW) and cosmic microwave background (CMB) experiments, as we discuss in the upcoming sections. While earlier works \cite{Bernal:2021yyb, Mazde:2022sdx, Borah:2024qyo} considered vacuum misalignment of the axion field in a PBH dominated Universe, our present work differs from these earlier works in two aspects namely, (i) inclusion of axion kinetic misalignment and (ii) inclusion of memory-burden (MB) effect of PBH. The memory-burden effect was pointed out in recent works \cite{Dvali:2018xpy, Dvali:2018ytn, Dvali:2024hsb} to include the backreaction effects of emitted particles on the black hole (BH) itself which was not taken into account in Hawking's original semi-classical (SC) approach to BH evaporation. We show the change in axion parameter space by comparing two different misalignment mechanisms in a PBH dominated Universe with two different types of evaporation namely, the one with MB effect and the other with standard SC approximation. We also show the corresponding detection prospects at axion detection, GW and CMB experiments.

This paper is organised as follows. In section \ref{sec2}, we summarise the vacuum and kinetic misalignment mechanisms of axion production in standard cosmology. In section \ref{sec3}, we discuss the basics of primordial black holes and their evaporation in semi-classical and memory-burdened regimes. In section \ref{sec4}, we discuss axion misalignments in PBH dominated Universe considering SC approximation as well as MB effects. In section \ref{sec5} we discuss the detection prospects and finally conclude in section \ref{sec6}.

\section{Axion misalignment in standard cosmology}
\label{sec2}
In a typical QCD axion model, the SM is extended by a Peccei-Quinn global symmetry $U(1)_{\rm PQ}$ \cite{Peccei:1977hh, Peccei:1977ur} with a singlet scalar $\sigma$ charged under it. A recent review of QCD axion models can be found in \cite{DiLuzio:2020wdo}. For simplicity, we consider the SM fields to be neutral under $U(1)_{\rm PQ}$ while $\sigma$ and a heavy vector-like quark $Q_{R}\,(Q_{L})$ have $U(1)_{\rm PQ}$ charges $1, -1/2\, (+1/2)$ respectively. This is similar to a KSVZ type model \cite{Kim:1979if, Shifman:1979if} mentioned earlier. The real part of the singlet scalar field acquires a vacuum expectation value (VEV) $v_{\rm PQ}=f_a$ such that it can be parametrised as $\sigma \equiv \dfrac{v_{\rm PQ} + \rho}{\sqrt{2}}\, e^{i {a}/{f_{a}}}$. The relevant part of the PQ invariant Yukawa Lagrangian can be written as
\begin{equation}\label{yukawalag}
    \mathcal{L_Y} \, \supset -  y \overline{Q}_L \sigma Q_R \, + {\rm h.c.}
\end{equation}
The scalar potential of the model is given by 
\begin{equation}
    V(H, \sigma) \, = \, \lambda_H \left(H^\dagger H \, - \, \dfrac{v^2}{2}\right)^2 \, + \, \lambda_\sigma \left(|{\sigma}|^2\, - \, \dfrac{v_{\rm PQ}^2}{2}\right)^2 \, +\, \lambda_{H \sigma} \left(H^\dagger H \, - \, \dfrac{v^2}{2}\right) \, \left(|{\sigma}|^2\, - \, \dfrac{v_{\rm PQ}^2}{2}\right)\, 
\end{equation}
with $v $ being the VEV of the neutral component of the SM Higgs $H$. After PQ symmetry breaking the terms in $\mathcal{L_Y}$ give rise to
\begin{eqnarray}
     y \bar{Q}_L \sigma Q_R \, \rightarrow \,  \frac{y}{\sqrt{2}} \rho \bar{Q}_L Q_R  e^{i {a}/{f_{a}}} \, + \, \frac{y}{\sqrt{2}} v_{PQ} \bar{Q}_L Q_R  e^{i {a}/{f_{a}}}.
\end{eqnarray}
The phase part can be absorbed by the transformation $ Q_{R} \, \rightarrow \, e^{-i\frac{a}{2 f_a}} Q_R$. As the chiral transformation on $Q$ is anomalous under QCD, this gives the term $\frac{g^2_s}{32 \pi^2} \frac{a}{f_a} G_{\mu \nu} \Tilde{G}^{\mu \nu}$ with $G_{\mu \nu} \, (\Tilde{G}^{\mu \nu})$ being (dual) field strength tensor of $SU(3)_c$ in QCD. Simultaneously, from the kinetic term $\overline{Q}i\gamma^{\mu}\partial_{\mu} Q$, after transformation one gets the term $-\frac{\partial_{\mu} a}{2 f_{a}}\overline{Q}\gamma^{\mu} \gamma_{5} Q$. Now using the Dirac equation and the fact that the total derivative is zero at the boundary, we get 
\begin{eqnarray}
    -\frac{\partial_{\mu} a}{2 f_{a}}\overline{Q}\gamma^{\mu} \gamma_{5} Q \, = \,  i \frac{M_Q}{f_{a}} a \overline{Q} \gamma_{5} Q 
\end{eqnarray}
which define axion couplings to heavy quark $Q$. For QCD axions, the zero temperature mass ($T\leq T_{\rm QCD} \simeq$ 160 MeV) is related to
PQ symmetry breaking scale, $f_{a}$ as
\begin{align}
    m_{a} \simeq 5.7 \left(\frac{10^{12}~\rm GeV}{f_{a}}\right) \rm \mu eV. 
\end{align}
Above $T>T_{\rm QCD}$, the temperature-dependent axion mass is given as 
\begin{align}
    m_{a} (T) = m_{a} \left(\frac{T_{\rm QCD}}{T}\right)^{4}.
\end{align}
As indicated by lattice simulation, the power of 4 is not precise \cite{DiLuzio:2020wdo}, but it does not affect our overall results significantly. 

For large decay constant $f_a$ as required from astrophysical bounds, the axion couplings to the SM particles are highly suppressed keeping thermal and non-thermal production of axions from the bath highly suppressed. However, in the ultra-light axion mass window, the axion field can be considered as a coherently oscillating scalar field. This can lead to production of cold axions within the purview of misalignment mechanisms which we briefly summarise below.

\subsection{Vacuum Misalignment}
The evolution of axion field $a$ in the early Universe can be written as
\begin{eqnarray}
    \Ddot{a} + 3\mathcal{H} \Dot{a} + \frac{1}{R^2(t)} \nabla ^2 a + \frac{\partial V(a, T)}{\partial a} = 0,
\label{axioneq}
\end{eqnarray}
where 
\begin{eqnarray}
    V(a, T) = f^2_{a} m^2_{a}(T) \left(1-\cos{\left(\frac{a}{f_{a}}\right)}\right)
\end{eqnarray}
and $R, \mathcal{H}$ denote the scale factor, Hubble expansion parameter respectively. The initial axion angle, $\theta = \frac{a}{f_{a}}$ is frozen in until the oscillation temperature of axion which can be estimated by comparing Hubble expansion rate to the temperature dependent axion mass 
\begin{eqnarray}
    \mathcal{H} (T_{\rm osc}) = 3m_{a} (T_{\rm osc}).
\label{Hem}
\end{eqnarray}
Initially $\theta$ is fixed at a constant value such that $ \theta_i \, \in \, (-\pi, \pi)$ and $\dot{\theta}_i \, = \,0$. As the temperature drops in an expanding Friedmann–Lema${\rm \hat{i}}$tre–Robertson–Walker (FLRW) Universe, at some point the condition given in Eq. \eqref{Hem} is satisfied leading to the onset of oscillations. This is when the axion starts behaving as pressure-less cold DM. The energy density at any given temperature is, \begin{equation}\label{energyden}
    \rho_a \, = \, \dfrac{1}{2} \, \dot{a}^2 + V(a)
\end{equation}
which can be found by solving Eq. \eqref{axioneq} numerically and the present axion abundance for different $f_a$ can accordingly be found by using the appropriate redshift factor. From the onset of oscillations temperature, axion starts behaving as matter, and from conservation of comoving number density, its number density at a later epoch can be written as 
\begin{align}
    n_{a}(T) = n_{a}(T_{\rm osc}) \frac{s(T)}{s(T_{\rm osc})}.
\end{align}
Here $s(T)$ denotes the comoving entropy density at a temperature $T$. The axion behaves as cold matter and can make all the observed dark matter abundance with total abundance 
\begin{align}
    \Omega_{a}h^2 = \frac{\rho_{a}(T_{0})}{\rho_{c}}h^2 = \frac{m_{a}(T_{0})}{\rho_{c}} \left(\frac{\rho_{a}(T_{\rm osc})}{m_{a}(T_{\rm osc})} \frac{s(T)}{s(T_{\rm osc})} \right) h^2.
\end{align}
 The initial value of the angle $\theta_i$ decides the amount of misalignment of the axion field initially and determines the final axion abundance. For example, taking the initial value $\theta_i = \pi/\sqrt{3} \simeq 1.81$, we get correct DM abundance $\Omega_{a}h^2=0.12$ for $f_a \, \sim \, 3.3 \times 10^{11} \,$ GeV. It should be noted that, we assume the Peccei-Quinn symmetry to be broken in post-inflationary era where the initial misalignment angle, $\theta_{i}= a_{i}/f_{a}$ takes the average value $\theta_{i} = \frac{\pi}{\sqrt{3}} \sim 1.81$. For symmetry breaking during inflation, one can have large isocurvature perturbations constrained by CMB observations, as summarised in appendix \ref{appen1}.

\subsection{Kinetic Misalignment}
In contrast with the vacuum misalignment where $\dot{\theta}_i =0$, kinetic misalignment \cite{Co:2019jts, Chang:2019tvx} heavily relies on $\dot{\theta}_i \, \neq \,0$, meaning a non-zero initial velocity for the axion field. Kinetic misalignment takes place only when kinetic energy $K= \dot{\theta}^2 (T) f_a^2/2$ is greater than the potential energy $V (a, T)$ at the conventional oscillation temperature $T \, = \, T_{\rm osc}$. In such a case, the axion field can overcome the potential barrier leading to change in the misalignment angle at a rate $\dot{\theta}$. This rolling of the axion field stops when its kinetic energy redshifts to the height of the potential barrier resulting in the axion being trapped in a potential minimum where it starts oscillating at a temperature $T' < T_{\rm osc}$. We can again solve for the axion evolution given by Eq.  \eqref{axioneq} numerically and get energy density given by Eq. \eqref{energyden}. The dark matter abundance thus obtained can be parameterised as \cite{Co:2019jts}
\begin{equation}\label{KMabun}
    \Omega_{a} h^2 \approx 0.12 \left(\dfrac{10^9}{f_a}\right) \, \left( \dfrac{Y_\theta}{40}\right).
\end{equation}
Here $Y_\theta \equiv \,\dot{\theta}f_a^2/s$, with $s$ being the entropy density, is the comoving density. Eq. \eqref{KMabun} is valid only when $Y_\theta \geq Y_{\rm crit}$. $Y_{\rm crit}$ is the critical density such that for $Y_\theta \leq Y_{\rm crit}$, Hubble friction will dampen the velocity $\dot{\theta}$ to zero before the conventional oscillation temperature $T \, = \, T_{\rm osc}$ effectively taking us to the ballpark of vacuum misalignment. Due to the delay in the oscillation temperature, the axion parameter space consistent with DM relic criteria significantly changes in kinetic misalignment scenario, as we discuss in the upcoming sections.

\section{Primordial black holes}
\label{sec3}
Primordial black holes, as the name suggests, are the black holes without stellar origin but produced from collapse of primordial over-densities in the early Universe. Originally proposed by Zeldovich \cite{Zeldovich:1967lct} and later by Hawking \cite{Hawking:1974rv, Hawking:1975vcx}, PBH\footnote{A comprehensive recent review of PBH can be found in \cite{Carr:2020gox}.} can have very interesting cosmological consequences \cite{Chapline:1975ojl, Carr:1976zz}. While PBH can be formed in a variety of ways like, from inflationary perturbations \cite{Hawking:1971ei, Carr:1974nx, Wang:2019kaf, Byrnes:2021jka, Braglia:2022phb}, first-order phase transition (FOPT) \cite{Crawford:1982yz, Hawking:1982ga, Moss:1994iq, Kodama:1982sf}, the collapse of topological defects \cite{Hawking:1987bn, Deng:2016vzb} and so on, we remain agnostic about such origin of PBH and assume them to form in the radiation dominated Universe at temperature $T_{\rm in}$. We also consider the PBH to be of Schwarzschild type having a monochromatic mass function having with initial mass $M_{\rm in}$ and initial energy fraction
\begin{eqnarray}
    \beta \equiv \frac{\rho_{\rm BH} (T_{\rm in})}{\rho_{R}(T_{\rm in})},
\end{eqnarray}
where $\rho_{\rm BH}$, $\rho_{\rm R}$ are the PBH and radiation energy densities respectively. The initial mass of PBH from gravitational collapse is
typically close to the mass enclosed by the post-inflationary particle horizon given by
\begin{equation}
     M_{\rm in}=\gamma \frac{4\,\pi\,}{3\,\mathcal{H}\left(T_\text{in}\right)^{3}}\,\rho_\text{R}\left(T_\text{in}\right)\, ,
\end{equation}
where $\gamma \simeq 0.2$ is an uncertainty parameter related to PBH formation \cite{Carr:1974nx}.  Given that PBH forms during early radiation dominated era, the epoch of formation can be written as 
\begin{eqnarray}
    t_{\rm in} = \frac{M_{\rm in}}{8\, \pi \gamma M_{P}^2},
\end{eqnarray}
with $M_{P}$ denoting the reduced Planck mass. Using the time-temperature relation in a radiation dominated FLRW Universe, we can then find PBH formation temperature as
\begin{equation}
T_\text{in}=\Biggl(\frac{1440\,\gamma^2}{g_\star\left(T_\text{in}\right)}\Biggr)^{1/4}\,\sqrt{\frac{M_{P}}{M_\text{in}}}\,M_{P}\,.
\label{eq:pbh-in}
\end{equation}
The instantaneous temperature and entropy associated with a black hole of mass $M_{\rm BH}$ are given as 
\begin{eqnarray}
    T_{\rm BH} &=& \frac{M_{P}^2}{M_{\rm BH}}, \,\,\,\,
    S = \frac{1}{2} \left(\frac{M_{\rm BH}}{M_{ P}}\right)^2 = \frac{1}{2} \left(\frac{M_{P}}{T_{\rm BH}}\right)^2, 
    \label{eq:entropy}
\end{eqnarray}
respectively. After formation, PBH can evaporate by emitting Hawking radiation \cite{Hawking:1974rv, Hawking:1975vcx} which we summarise below using Hawking's semi-classical approximation and recently formulated memory-burden effect.

\subsection{PBH evaporation in semi-classical approximation}
In the semi-classical approximation of Hawking, the PBH mass loss rate is given by \cite{MacGibbon:1991tj}
\begin{eqnarray}
    \frac{dM_{\rm BH}}{dt} = - \epsilon \frac{M^4_{P}}{M^2_{\rm BH}},
    \label{eq:massloss}
\end{eqnarray}
where 
\begin{equation}
\epsilon = \frac{27}{4} \frac{\pi g_{*, H}(T_{\rm BH})}{480}, \,\, g_{*, H}(T_{\rm BH}) = \sum_i \omega_i g_{i,H}, \,\,g_{i,H}=
    \begin{cases}
        1.82
        &\text{for }s_i=0\,,\\
        1.0
        &\text{for }s_i=1/2\,,\\
        0.41
        &\text{for }s_i=1\,,\\
        0.05
        &\text{for }s_i=2\,,\\
    \end{cases}
    \label{eqn:gsh}
\end{equation}
with $\omega_i=2s_i+1$ for massive particles of spin $s_i$, $\omega_i=2$ for massless species with $s_i>0$, and $\omega_i=1$ for spinless species $s_i=0$. Integrating Eq. \eqref{eq:massloss}, gives PBH mass at any epoch after its formation as
\begin{eqnarray} \label{eq:MBH_time_SC}
    M_{\rm BH}(t) = M_{\rm in} \left(1-\frac{3\,\epsilon\,M_{P}^4}{M_{\rm in}^3}(t-t_{\rm in})\right)^{1/3} \equiv  M_{\rm in} \left(1- \Gamma^0_{\rm BH}(t-t_{\rm in})\right)^{1/3},
\end{eqnarray}
where, $\Gamma^0_{\rm BH}=\frac{3\,\epsilon\,M_{P}^4}{M_{\rm in}^3}$ is similar to decay width in SC approximation. Assuming the validity of the SC approximation till complete evaporation, the PBH lifetime $t^0_{\rm ev}\gg t_{\rm in}$ can be found as
\begin{eqnarray} \label{eq:t0ev}
    t^0_{\rm ev} = \frac{1}{\Gamma^0_{\rm BH}} = \frac{M_{\rm in}^3}{3\,\epsilon\,M_{P}^4}.
\end{eqnarray}
The corresponding evaporation temperature can then be computed taking into account $\mathcal{H}(T_\text{ev})\sim\frac{1}{(t^0_{\rm ev})^2}\sim\rho_\text{R}(T_\text{ev})$ as
\begin{equation}
T_\text{ev}\equiv\Bigl(\frac{90\,M_P^2}{4\,\pi^2\,g_\star\left(T_\text{ev}\right)\, (t^0_{\rm ev})^2}\Bigr)^{1/4}\,.
\label{eq:pbh-Tev}
\end{equation}
However, if the PBH component dominates the total energy density of the Universe at some epoch, the SM temperature just after the complete evaporation of PBHs is $\overline{T}_\text{ev}=2/\sqrt{3}\,T_\text{ev}$~\cite{Bernal:2020bjf}. PBH can dominate the energy density of the early Universe, if their initial energy density is greater than a critical value given by
\begin{align}
    \beta \geq \beta_{\text{c}} \simeq 2.5\times 10^{-14} \gamma^{-\frac{1}{2}} \left(\frac{M_{\rm in}}{10^8 \text{g}}\right)^{-1}\,. \label{eq:betacr}
\end{align}

\subsection{PBH evaporation with memory-burden effect}
In the semi-classical approximation of Hawking \cite{Hawking:1974rv, Hawking:1975vcx}, the backreaction of the emitted particles on the black hole itself was ignored. As pointed out recently \cite{Dvali:2018xpy, Dvali:2018ytn, Dvali:2024hsb}, such backreaction, referred to as memory-burden effect, can slow down the rate of PBH evaporation specially after the energy of emitted particles become comparable to that of PBH. The enhanced PBH lifetime in memory-burdened regime can have interesting phenomenological consequences for dark matter, gravitational waves, baryon asymmetry of the Universe and high energy astroparticle physics as have been discussed in several recent works \cite{Dvali:2021byy, Balaji:2024hpu, Barman:2024iht, Bhaumik:2024qzd, Barman:2024ufm, Kohri:2024qpd, Jiang:2024aju, Zantedeschi:2024ram, Chianese:2024rsn, Barker:2024mpz, Alexandre:2024nuo, Thoss:2024hsr, Haque:2024eyh, Borah:2024bcr, Loc:2024qbz, Basumatary:2024uwo, Athron:2024fcj, Barman:2024kfj}.

\begin{figure}
\centering
  \includegraphics[width=0.5\linewidth]{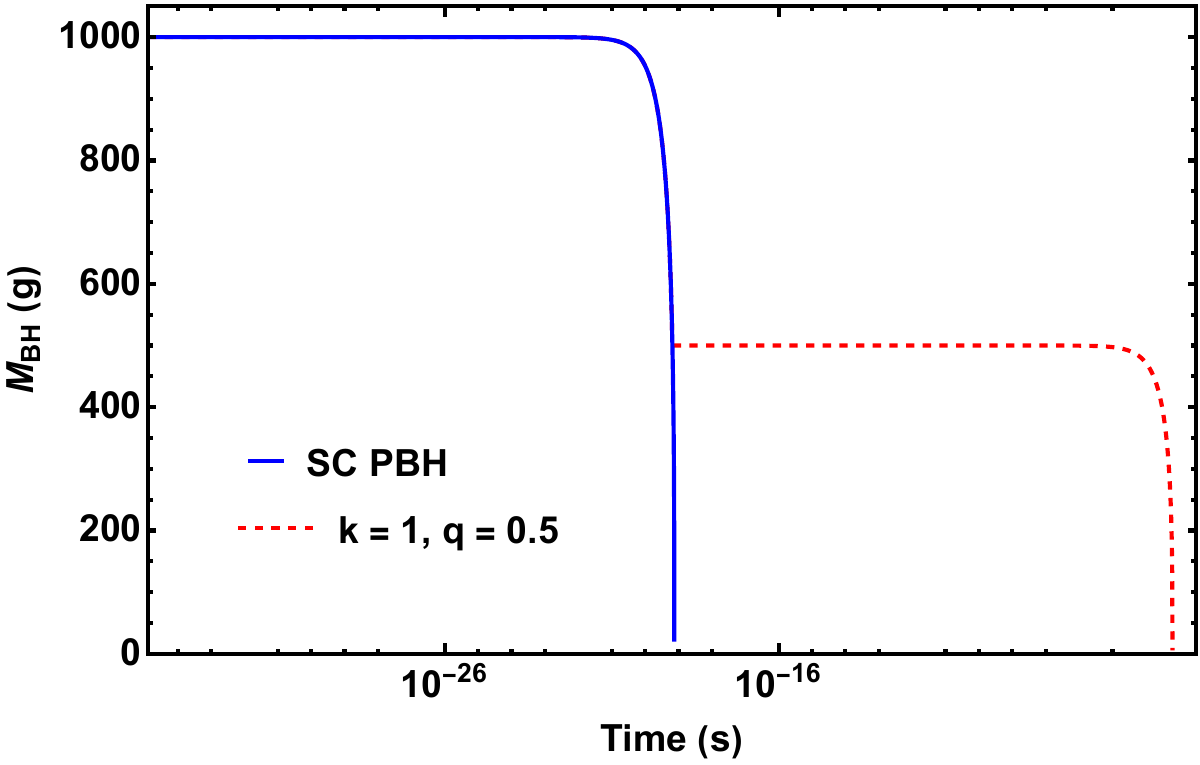}
    \caption{Evolution of PBH mass in SC (blue solid line) and MB regime with $k=1, q=0.5$ (red dashed line).}
    \label{fig1}
\end{figure}

In such a scenario, a black hole is assumed to evaporate in a semi-classical manner till its instantaneous mass becomes a fraction of its initial mass $M_{\rm BH}=q\,M_{\rm in}$, where $0<q<1$. Thus, in the PBH mass range $(q M_{\rm in}, 0)$, quantum memory effects dominate to alter the evaporation rate followed in the semi-classical regime. In this memory-burden regime, the PBH evaporation rate is given as
\begin{eqnarray}
     \frac{dM_{\rm BH}}{dt} = - \frac{\epsilon}{[S(M_{\rm BH})]^k} \frac{M^4_{P}}{M^2_{\rm BH}},
     \label{eq:massloss2}
\end{eqnarray}
where $S$ is the black hole entropy defined in Eq. \eqref{eq:entropy}. Integrating this from an initial mass $q M_{\rm in}$ at $t=t_{q}$ to a later epoch $t$, we get
\begin{eqnarray} \label{eq:MBH_time_MB}
    M_{\rm BH} (t) = q M_{\rm in} \left(1- \Gamma^k_{\rm BH}(t-t_{q})\right)^{1/(3+2k)},
\end{eqnarray}
where 
\begin{eqnarray}
    \Gamma^{k}_{\rm BH} \equiv 2^k (3+2k) \, \epsilon \, M_{P} \left(\frac{M_{P}}{qM_{\rm in}}\right)^{3+2k}
\end{eqnarray}
is the associated decay width for the MB regime. The total lifetime of a PBH then can be written as 
\begin{eqnarray}
    t^k_{\rm ev} = t_{q} + \frac{1}{\Gamma^{k}_{\rm BH}} = \frac{1-q^3}{\Gamma^0_{\rm BH}}  + \frac{1}{\Gamma^{k}_{\rm BH}}. 
\end{eqnarray}
The critical value of $\beta$, denoted as $\beta_{\rm c}$ can be obtained as 
\begin{eqnarray}
    \beta_{\rm c} = \left(\frac{(3+2k)2^{k}\epsilon}{8\, q^3 \pi \gamma}\right)^{1/2} \left(\frac{M_{P}}{q M_{\rm in}}\right)^{1+k},
\end{eqnarray}
such that $\beta > \beta_{c}$ indicates an early matter domination era due to PBH. The evaporation temperature with MB effect can be found as 
\begin{equation}
    T_{\rm ev}= M_{P} \left(\frac{4}{3\alpha'}\right)^{1/4} \left(\frac{3\times2^{k}(3+2k)\epsilon \left(\frac{M_{P}}{M_{\rm in}}\right)^{3+2k}}{3\times q^{3+2k}+(1-q^3)2^{k}(3+2k)\left(\frac{M_{P}}{M_{\rm in}}\right)^{2k}}\right)^{1/2}.
\end{equation}
Here $\alpha' = \frac{\pi^2}{30} \, g_{*} (T_{\rm ev})$, with $g_{*} (T_{\rm ev})$ being the relativistic degrees of freedom associated with SM bath at $T=T_{\rm ev}$. For either $q\to0$ or $k\to 0$, the above expression reduces to standard semi-classical expression. The details of these derivations can be found in \cite{Borah:2024bcr}. Fig. \ref{fig1} shows the evaporation rate comparison of SC and MB regimes clearly indicating the slowing down of evaporation rate in the latter.

\begin{figure}
\centering
  \includegraphics[width=0.49\linewidth]{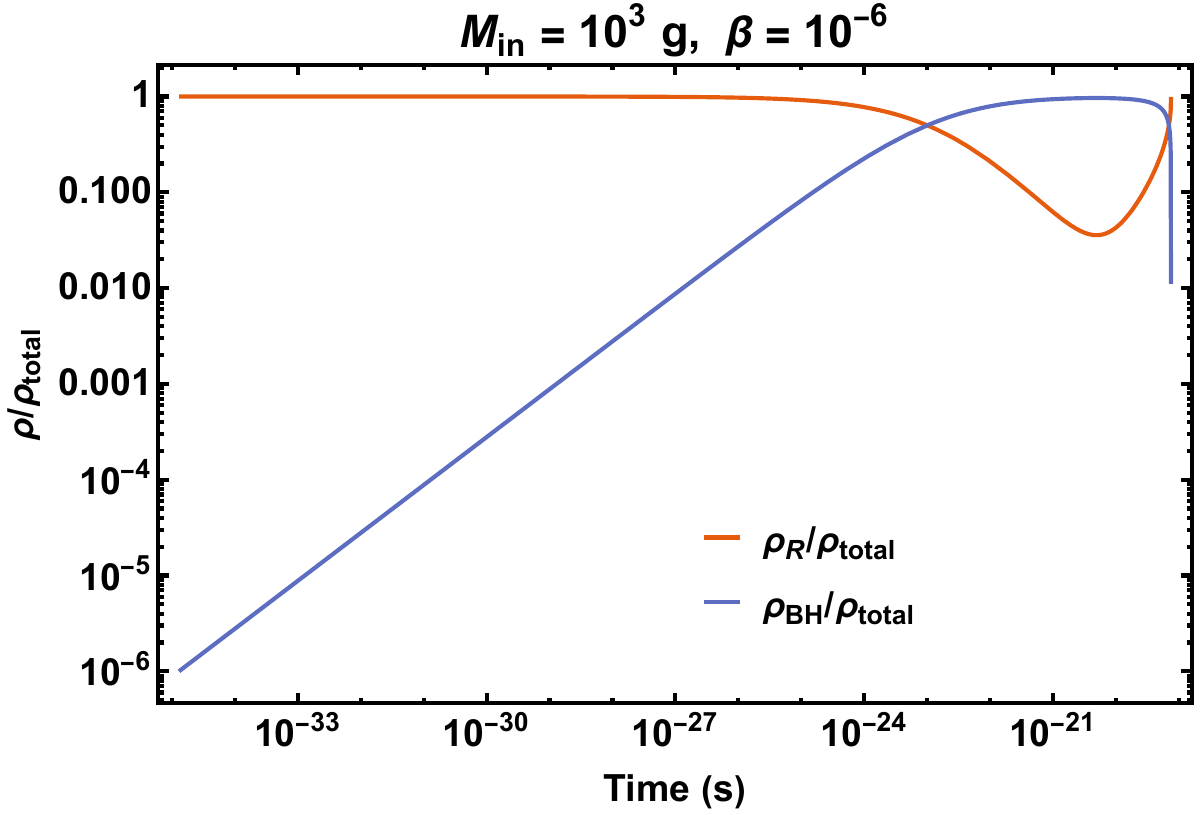}
  \includegraphics[width=0.49\linewidth]{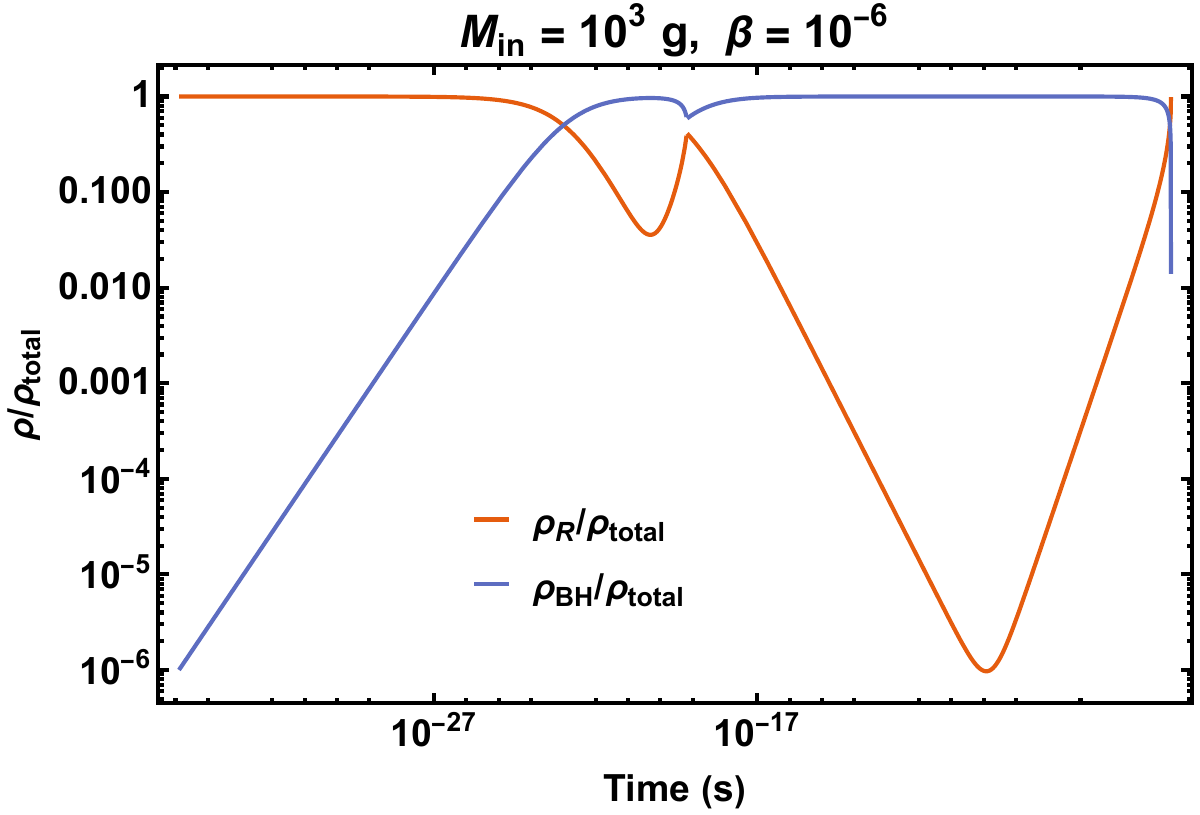}
    \caption{Evolution of radiation and PBH energy densities in SC (left panel) and MB regime with $k=1, q=0.5$ (right panel).}
    \label{fig2}
\end{figure}

%\begin{figure}
%\centering
%  \includegraphics[width=0.49\linewidth]{Figure/timetemp.pdf}
%  \includegraphics[width=0.49\linewidth]{Figure/masshub.pdf}
%    \caption{Left panel: temperature evolution with PBH in SC regime and without PBH. Right panel: Evolution of temperature dependent axion mass and Hubble rate with PBH in SC regime and without PBH.}
%    \label{fig2a}
%\end{figure}

%\begin{figure}
%\centering
%   \includegraphics[width=0.49\linewidth]{Figure/vacM_time.pdf}
%      \includegraphics[width=0.49\linewidth]{Figure/betavac_time.pdf}
%\includegraphics[width=0.49\linewidth]{Figure/kvac_time.pdf}
%\includegraphics[width=0.49\linewidth]{Figure/qvac_time.pdf}
%    \caption{Top panel: evolution of $\theta$ for vacuum misalignment in presence of PBH without memory burden effect. Bottom panel: evolution of $\theta$ for vacuum misalignment in presence of PBH with memory burden effect considering different value of $(k, q)$.}
%    \label{fig3}
%\end{figure}

\section{Axion misalignment in presence of PBH}
\label{sec4}
A PBH dominated era in the early Universe can modify the axion abundance produced via misalignment mechanism either by changing the oscillation temperature $T_{\rm osc}$ or by diluting the axion abundance via late entropy injection \cite{Bernal:2021yyb, Mazde:2022sdx}. Both of these effects can be incorporated by numerically solving the relevant evolution equations for axion, radiation and PBH simultaneously.

The Boltzmann equations for PBH and radiation energy densities can be written as
\begin{eqnarray}
\frac{d \rho_{\rm BH}}{dt} + 3 \mathcal{H} \rho_{\rm BH} = \frac{1}{M_{\rm BH}} \frac{d M_{\rm BH}}{dt} \rho_{\rm BH} \label{eqn:beq}\\
\frac{d \rho_{R}}{dt} + 4 \mathcal{H} \rho_{R} = - \frac{1}{M_{\rm BH}} \frac{d M_{\rm BH}}{dt} \rho_{\rm BH}\label{eqn:beq1},
\end{eqnarray}
where Hubble expansion rate is 
\begin{eqnarray}\label{eqn:beq2}
    3 M_{\rm P} \mathcal{H}^2 =  \rho_{R} + \rho_{\rm BH}
\end{eqnarray}
and mass loss rate of PBH namely, $\frac{1}{M_{\rm BH}} \frac{d M_{\rm BH}}{dt}$ is given by Eq. \eqref{eq:massloss} and Eq. \eqref{eq:massloss2} in SC and MB regimes respectively. Fig. \ref{fig2} shows the evolution of radiation and PBH energy densities in SC and MB regimes of PBH evaporation for benchmark choices of PBH parameters. The two different periods of evaporation for PBH mass in the range $(M_{\rm in}, q M_{\rm in})$ and $(qM_{\rm in}, 0)$ are clearly visible on the right panel plot of Fig. \ref{fig2} showing the difference from the purely SC scenario shown on the left panel.

%The left panel of Fig. \ref{fig2a} shows the evolution of temperature with and without PBH for benchmark choices of PBH parameters in the SC regime. The right panel plot of the same figure shows the comparison of axion mass and Hubble rate with or without PBH which decides the axion oscillation temperature. 

If PBH dominate the energy density, leading to an early matter dominated era, their evaporation leads to entropy injection which can be tracked via
\begin{eqnarray}\label{eqn:beq3}
    \frac{ds}{dt} + 3 \mathcal{H} s = - \frac{1}{M_{\rm BH}} \frac{d M_{\rm BH}}{dt} \frac{\rho_{\rm BH}}{T}.
\end{eqnarray}
The above equations together with the axion evolution equation given by Eq. \eqref{axioneq} need to be solved simultaneously to find axion DM relic in the present Universe.  After finding the oscillation temperature $T_{\rm osc}$ in presence of PBH by tracking the axion evolution, the number density of axion at $T_{\rm osc}$ can be calculated as
\begin{eqnarray}
    n_{a}(T_{\rm osc}) = \frac{\rho_{a}(T_{\rm osc})}{m_{a}(T_{\rm osc})} = \frac{1}{{m_{a}(T_{\rm osc})}} \left(\frac{1}{2}\Dot{a}^2 + \frac{1}{2} |\nabla a|^2 + V(a)\right).
\end{eqnarray}
Due to entropy dilution, the equation for $n_{a}$ becomes 
\begin{eqnarray}
    \frac{1}{n_{a}} \frac{d(n_{a}/s)}{dt} = \frac{1}{M_{\rm BH}} \frac{d M_{\rm BH}}{dt} \frac{\rho_{\rm BH}}{T s^2}.
\end{eqnarray}
The above equation is then solved from $T \gtrsim T_{\rm osc}$ to a sufficiently lower temperature such that PBH evaporation becomes complete. 
\\
%\textcolor{Red}{If PBH evaporates after $T_{\rm osc}$ then entropy dilution comes into play and decreases the total abundance by decreasing the number density of axions. In order to calculate it, we solve for $a(t)$ by solving $\mathcal{H} = \dfrac{\dot{a}}{a}$. As total number of axions produced via misalignment mechanism remains constant, we use the relation $n_{a}(T_{\rm BBN}) = n_{a}(T_{\rm osc} )\, \dfrac{a^3_{\rm osc} }{a^3_{\rm BBN}}$. By the time of BBN, PBH has evaporated and entropy dilution has been accounted for by the change in the scale factor a(t).}
\begin{figure}
\centering
   \includegraphics[width=0.49\linewidth]{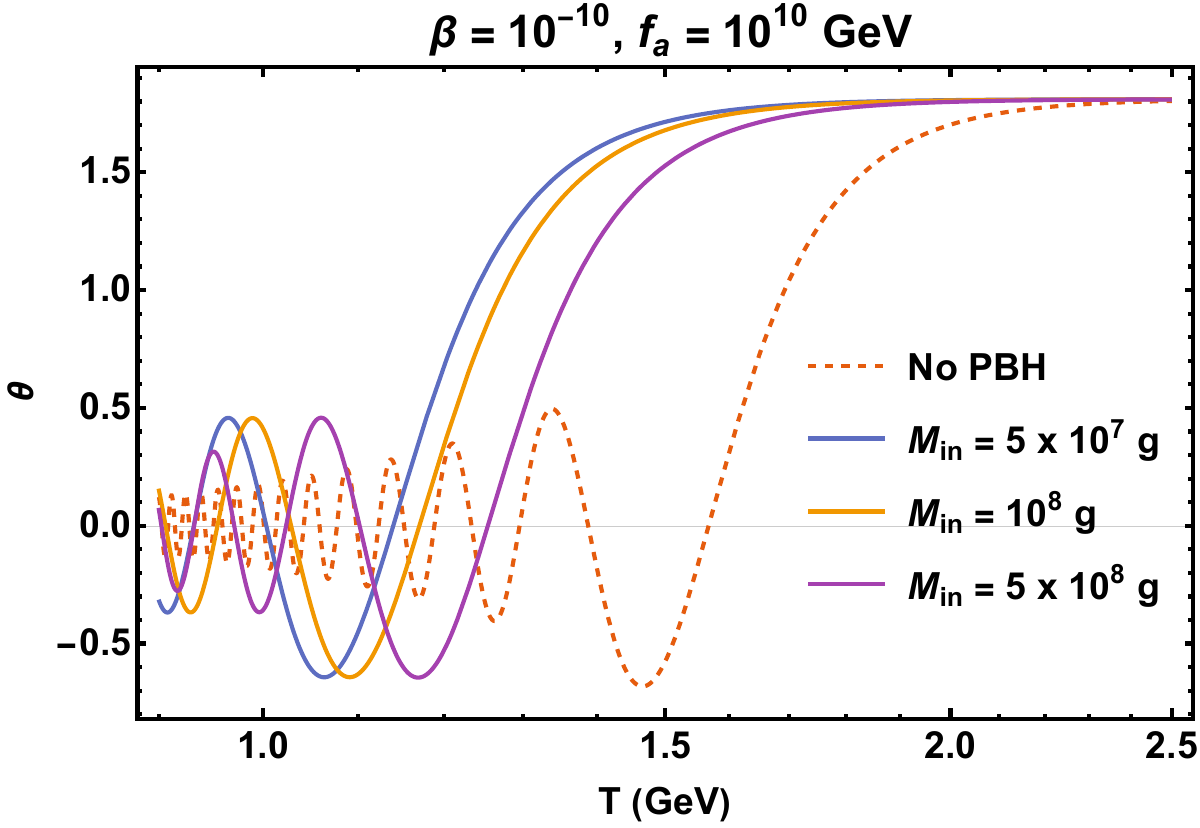}
      \includegraphics[width=0.49\linewidth]{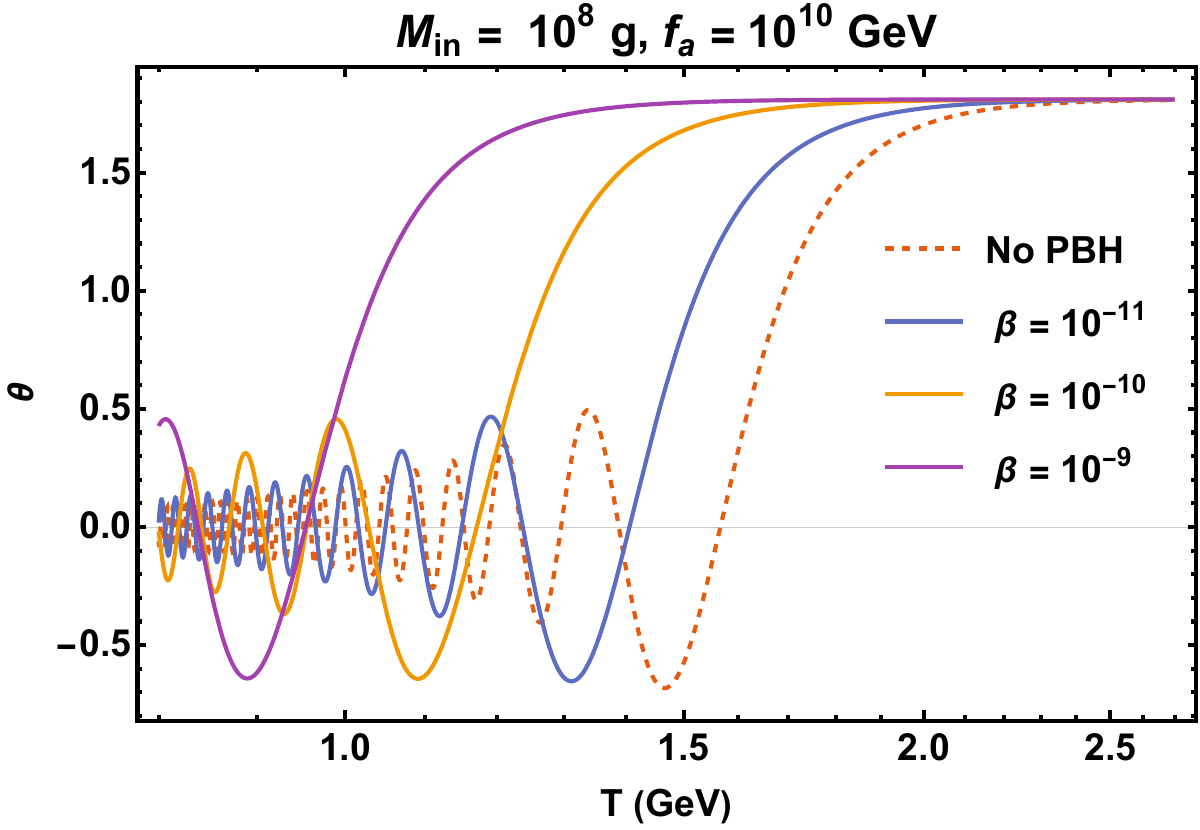}
\includegraphics[width=0.49\linewidth]{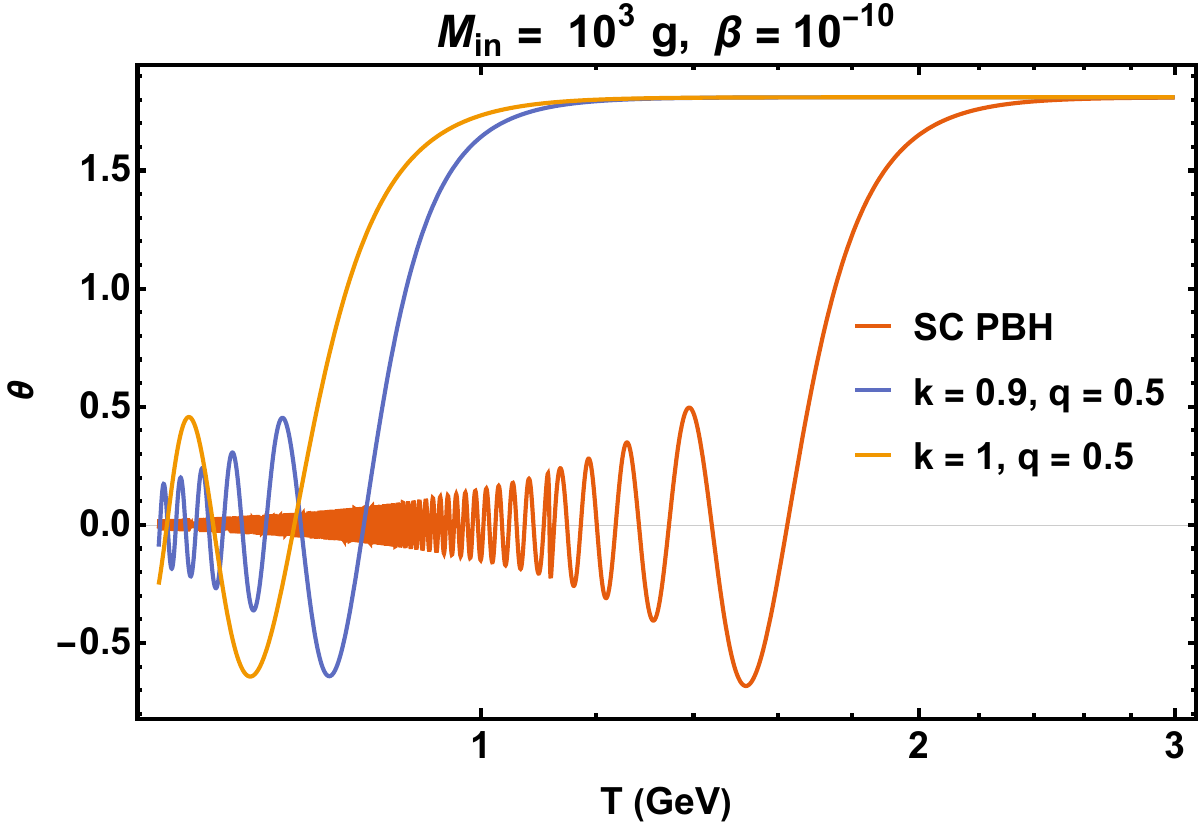}
\includegraphics[width=0.49\linewidth]{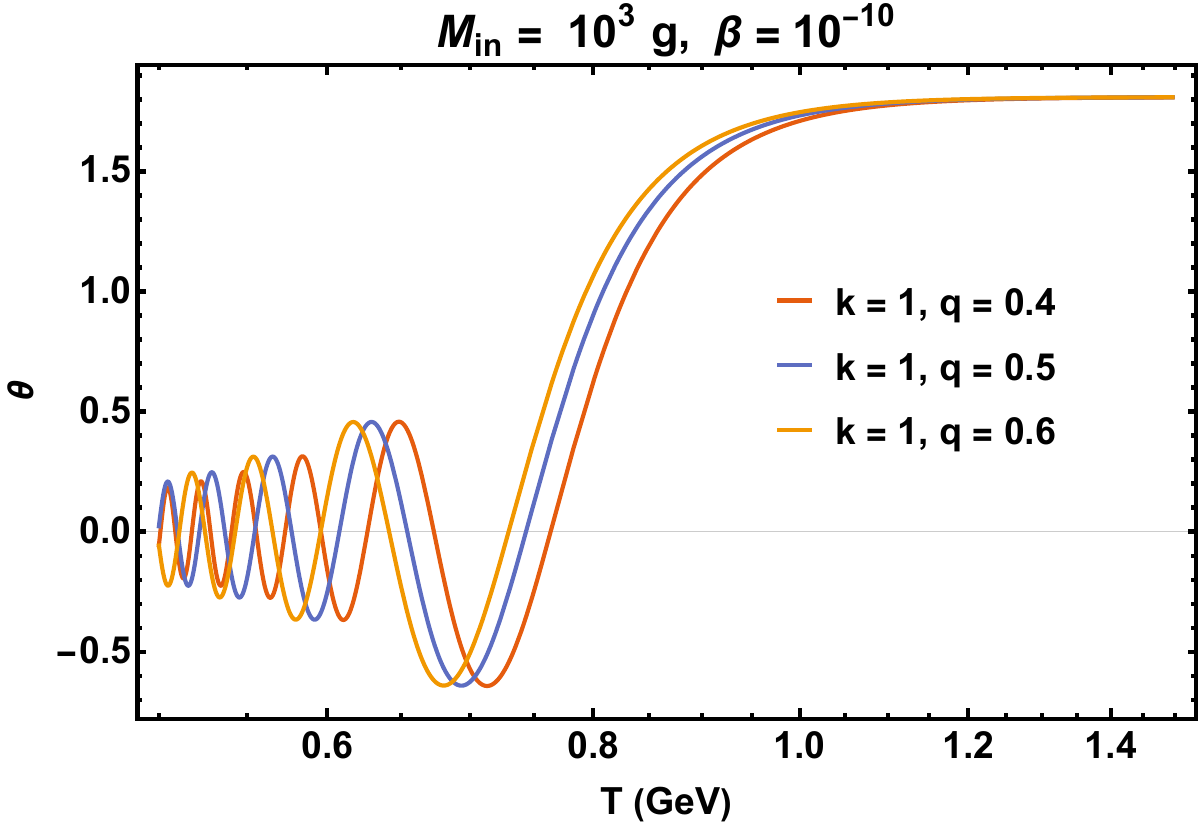}
    \caption{Top panel: evolution of $\theta$ for vacuum misalignment in presence of PBH without memory-burden effect. Bottom panel: evolution of $\theta$ for vacuum misalignment in presence of PBH with memory-burden effect considering different value of $(k, q)$.}
    \label{fig3}
\end{figure}
\begin{figure}
\centering
     \includegraphics[width=0.49\linewidth]{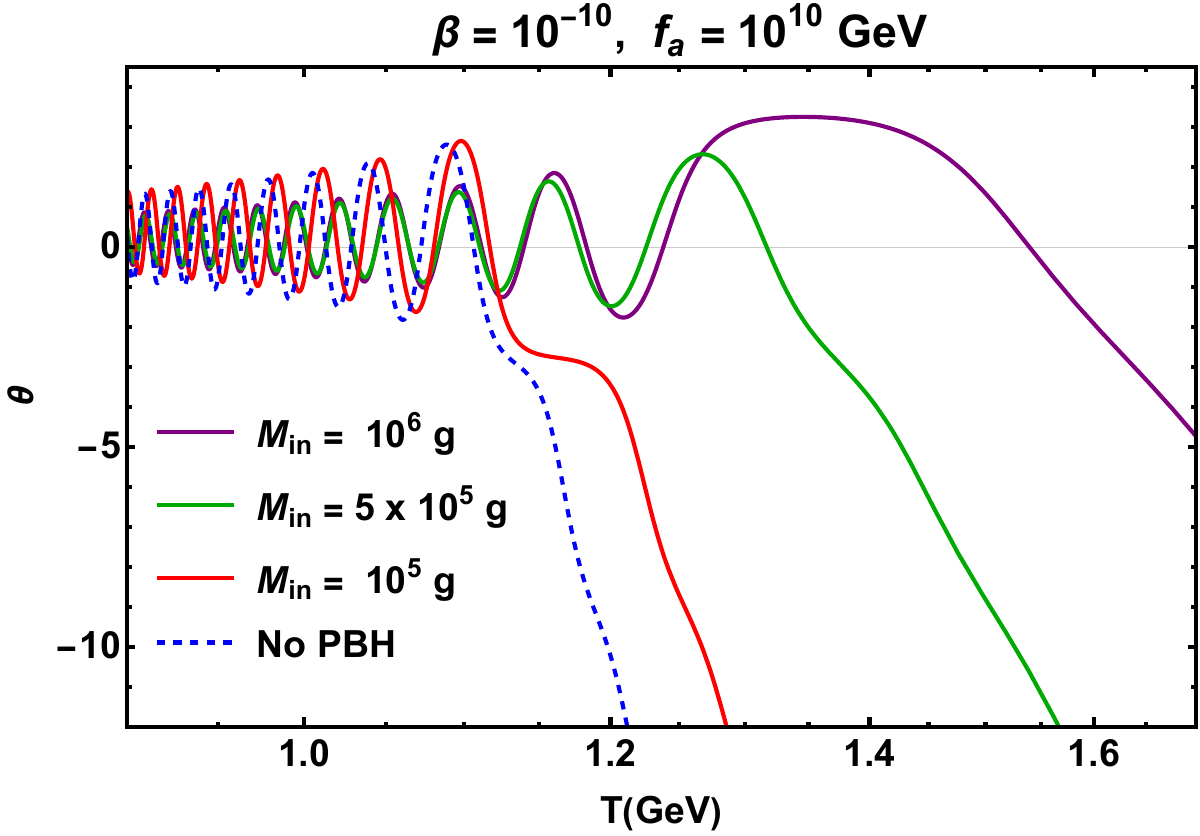}
          \includegraphics[width=0.49\linewidth]{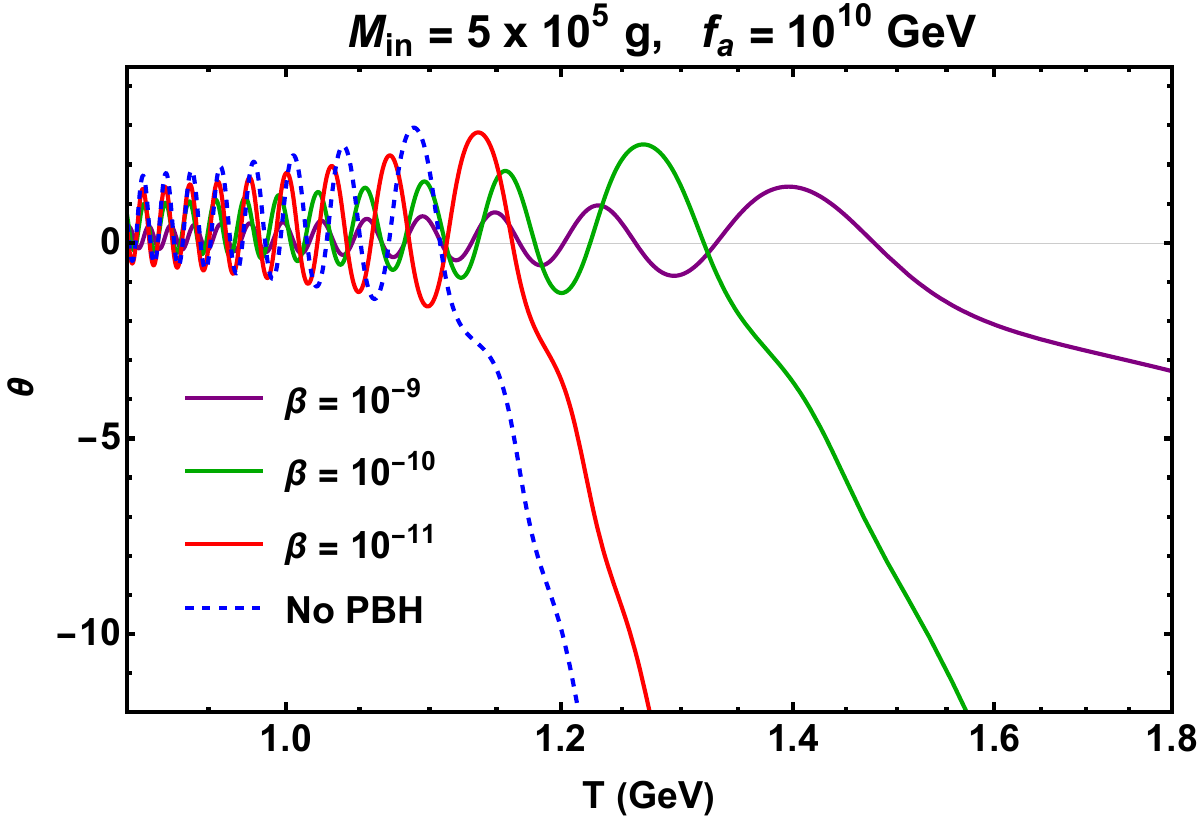}
             \includegraphics[width=0.49\linewidth]{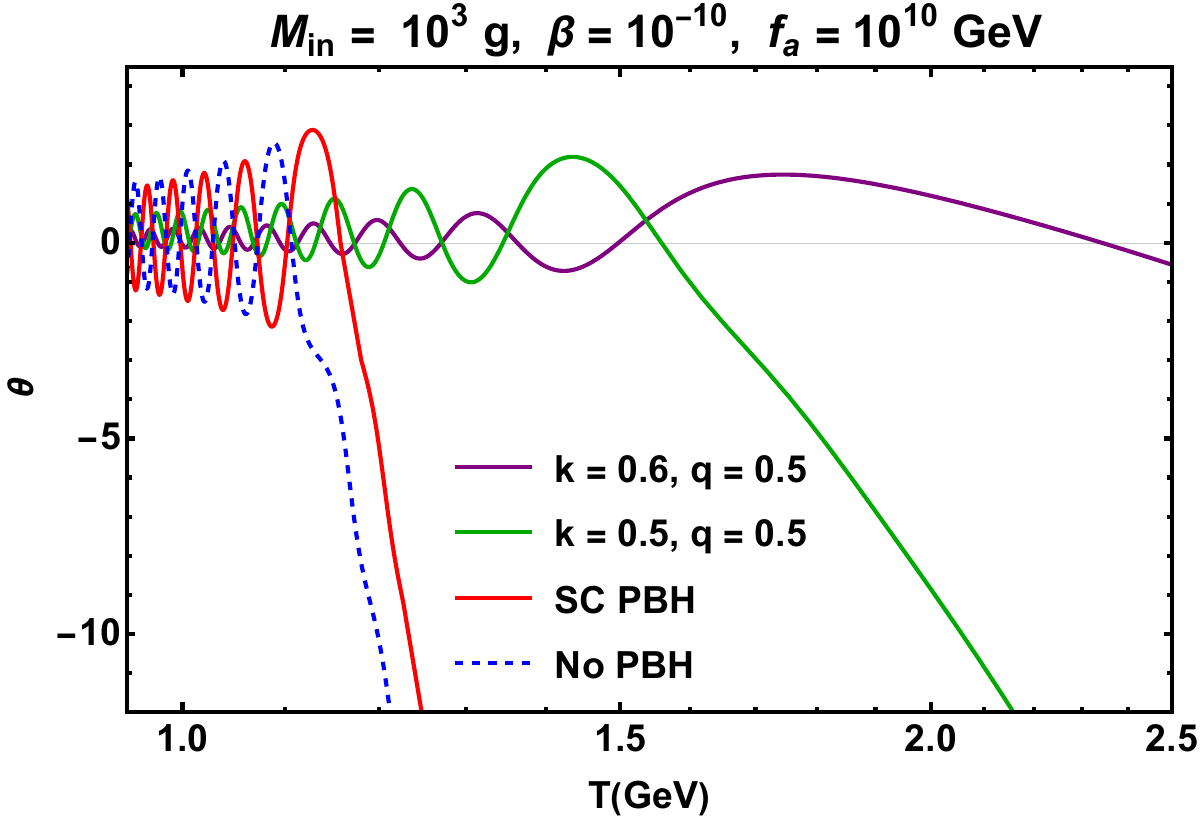}
     \includegraphics[width=0.49\linewidth]{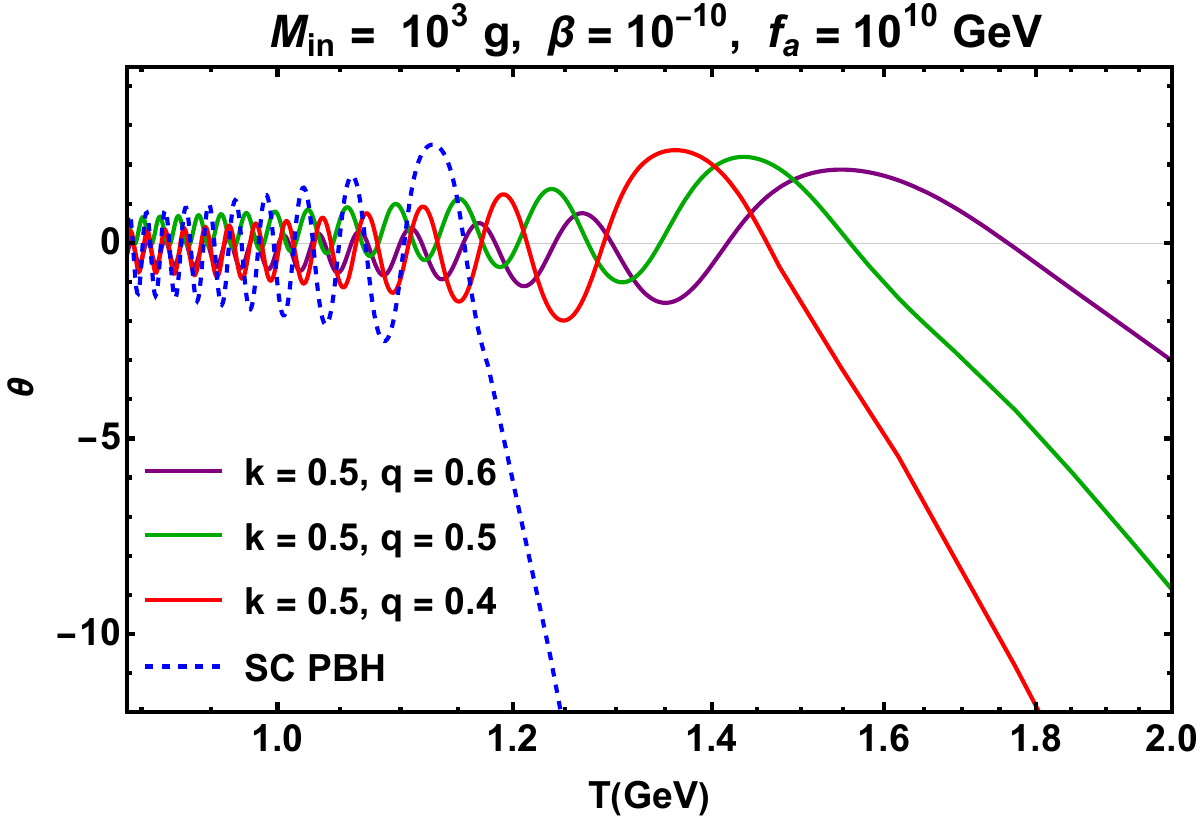}
    \caption{Top panel: evolution of $\theta$ for kinetic misalignment in presence of PBH without memory-burden effect. Bottom panel: evolution of $\theta$ for kinetic misalignment in presence of PBH with memory-burden effect considering different value of $(k, q)$. The initial velocity is $\dot{\theta}_i = 3 \times 10^{38} \, {\rm s}^{-1}$ at $T=f_a$.}
     %The Plot shows the evolution of $\theta$ without PBH and with varying PBH mass when PQ symmetry breaking Scale is $f_a = 10^{10}$ GeV. The evolution is shown for kinetic misalignment where the initial value of $\dot{\theta} = 10^{40} \, s^{-1}$ which is defined at T = $f_a$. As we can see $t_{osc}$ decreases with increase in mass in case of vacuum misalignment due to hubble rate being changed in presence of PBH for high mass PBH where oscillation takes place in PBH dominated universe or where PBH has some visible effect. An increased effect can be seen in kinetic misalignment even  for low mass PBH where oscillation does not occour in PBH  dominated universe but the difference is due to additional damping of $\dot{\theta}$ leading to higher oscillation temperature for higher masses. }
    \label{fig5}
\end{figure}

Fig. \ref{fig3} shows axion evolution in vacuum misalignment scenario for SC and MB regimes of PBH. Clearly, depending upon the parameters $(k, q)$ which quantify the MB effect, the axion oscillation can occur at different epochs. PBH domination increases the Hubble rate $\mathcal{H}$ leading to the onset of axion oscillations at a lower temperature compared to the one in standard FLRW cosmology. This effect increases as the contribution of PBH in the total energy density of Universe increases. Therefore, with increase in $\beta$ for fixed $M_{\rm in}$, oscillation temperature decreases as seen from the top right panel of Fig. \ref{fig3}. However, with fixed $\beta$ and increasing $M_{\rm in}$, the oscillation temperature increases as seen from the top left panel of Fig. \ref{fig3}. This is due to the decrease in PBH formation temperature with increasing $M_{\rm in}$, as described by Eq. \eqref{eq:pbh-in} that leads to a lower $\mathcal{H}$ at a fixed temperature for higher $M_{\rm in}$ when $\beta$ is held constant. As the lower panel plots of Fig. \ref{fig3} show, the increase in MB parameters $(k, q)$ delays the onset of axion oscillation by slowing down the evaporation of PBH. This will correspond to different allowed values of $\beta-M_{\rm in}-f_a$ required for correct axion DM abundance compared to the SC PBH scenario and can have different detection aspects for axion and GW experiments as well, which we discuss in upcoming sections.

Fig. \ref{fig5} shows the corresponding axion evolution for kinetic misalignment mechanism for fixed $f_a=10^{10}$ GeV while varying PBH parameters $(M_{\rm in}, \beta)$ together with $(k, q)$ in MB regime. Unlike the vacuum misalignment case where PBH domination delays the onset of axion oscillation decided by the equality of axion mass and Hubble according to Eq. \eqref{Hem}, here a larger Hubble expansion rate causes faster damping of velocity $\dot{\theta}$ forcing the oscillation to occur at a higher temperature compared to the one in standard scenario without PBH. Therefore, with increased PBH energy density dictated by larger $M_{\rm in}, \beta$ leads to higher oscillation temperature in SC regime shown in the top panel of Fig. \ref{fig5}. Similarly, with the increase in MB effect dictated by $(k, q)$, the oscillations occur at higher temperature, as shown in the bottom panel of Fig. \ref{fig5}. Clearly, the presence of PBH significantly changes the oscillation epoch or temperature compared to the scenario without PBH. Even in the presence of PBH, there exist sharp contrast between the SC and MB regimes as the oscillation epoch changes for different values of MB parameters $(k, q)$. In all these plots, the initial velocity is fixed to $\dot{\theta} = 3 \times 10^{38} \, {\rm s}^{-1}$ at $T=f_a$. 

%\textcolor{Blue}{\textbf{Kinetic - WRT TEMP:} Beyond the usual vacuum misalignment, in kinetic misalignment there comes an additional contribution of damping of $\dot{\theta}$ as the $\mathcal{H}$ also increases when PBH contributes to the total energy density of universe. So, with the increase in PBH mass, decrease in $\beta$ or with an increase in MB parameter k,q, the oscillation begans earlier at higher temperature.}\\
%\textcolor{Orange}{\textbf{Note:} In case of vacuum misalignment,  $T_{osc}|_{no -pbh} > T_{osc}|_{with- pbh}$, this is due to increase $\mathcal{H}$ at a given temperature, therefore the condition $3 \mathcal{H} \sim m_a$ is satisfied at a lower temperature.\\ In case of kinetic misalignment,  $T_{osc}|_{no-pbh} < T_{osc}|_{with-pbh}$ due to higher damping of $\dot{\theta}$ in presence of PBH.}

\begin{figure}
\centering
   \includegraphics[width=0.49\linewidth]{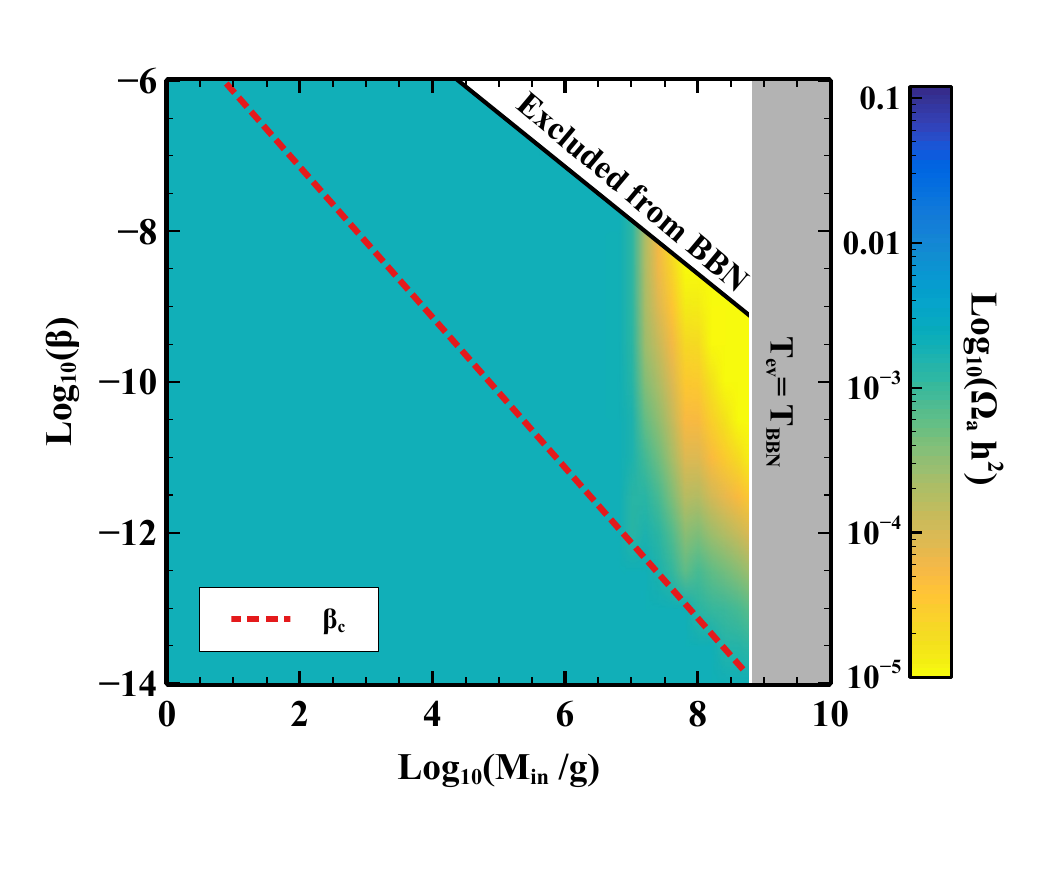}
      \includegraphics[width=0.49\linewidth]{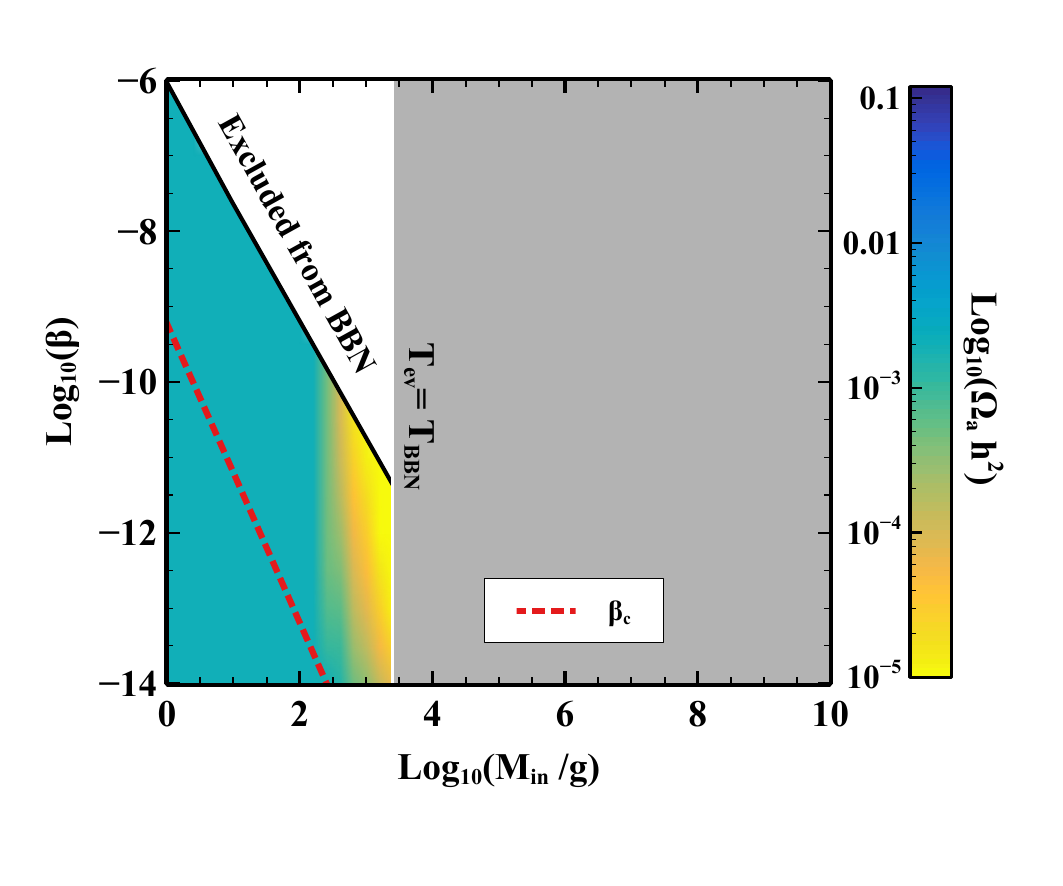}
    \caption{Variation of axion relic abundance from vacuum misalignment as a function of PBH parameters considering SC (left panel) and MB regime with $k=1, q=0.5$ (right panel). Here, $\theta_i \, = \pi/\sqrt{3}$ and $f_a=10^{10}$ GeV.}
    \label{fig6}
\end{figure}

\begin{figure}
\centering
   \includegraphics[width=0.49\linewidth]{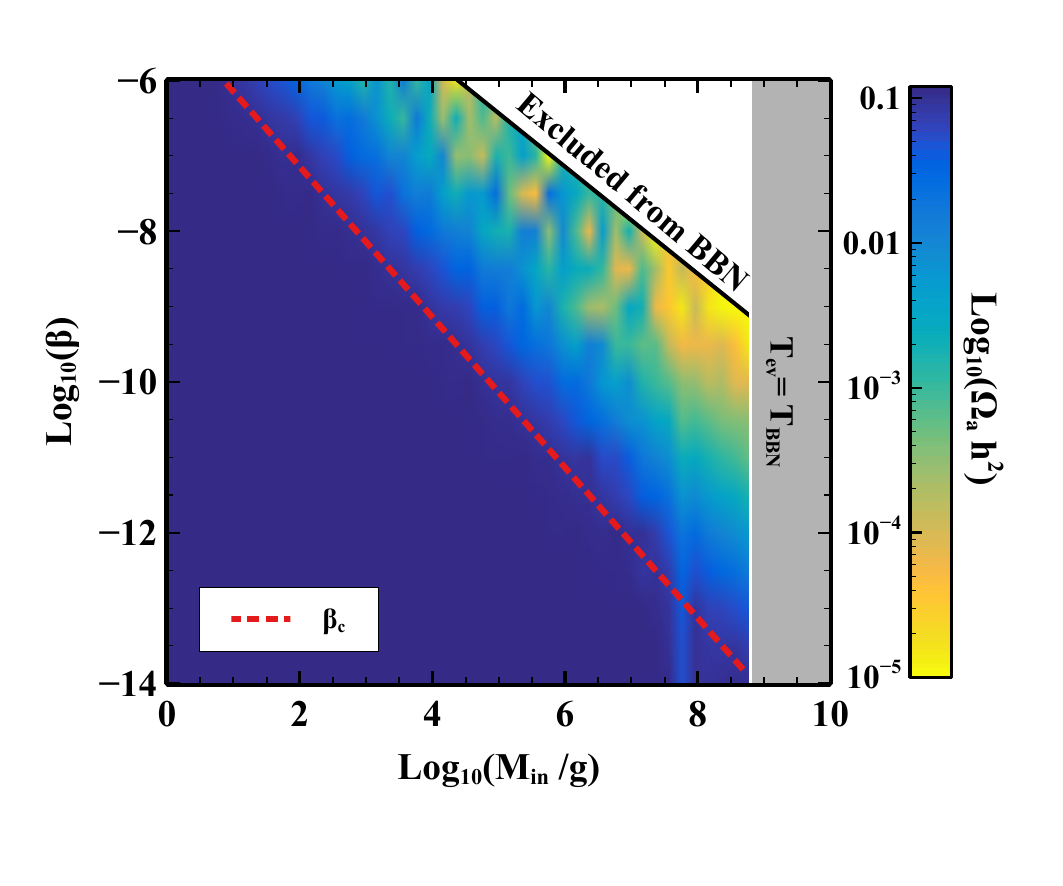}
      \includegraphics[width=0.49\linewidth]{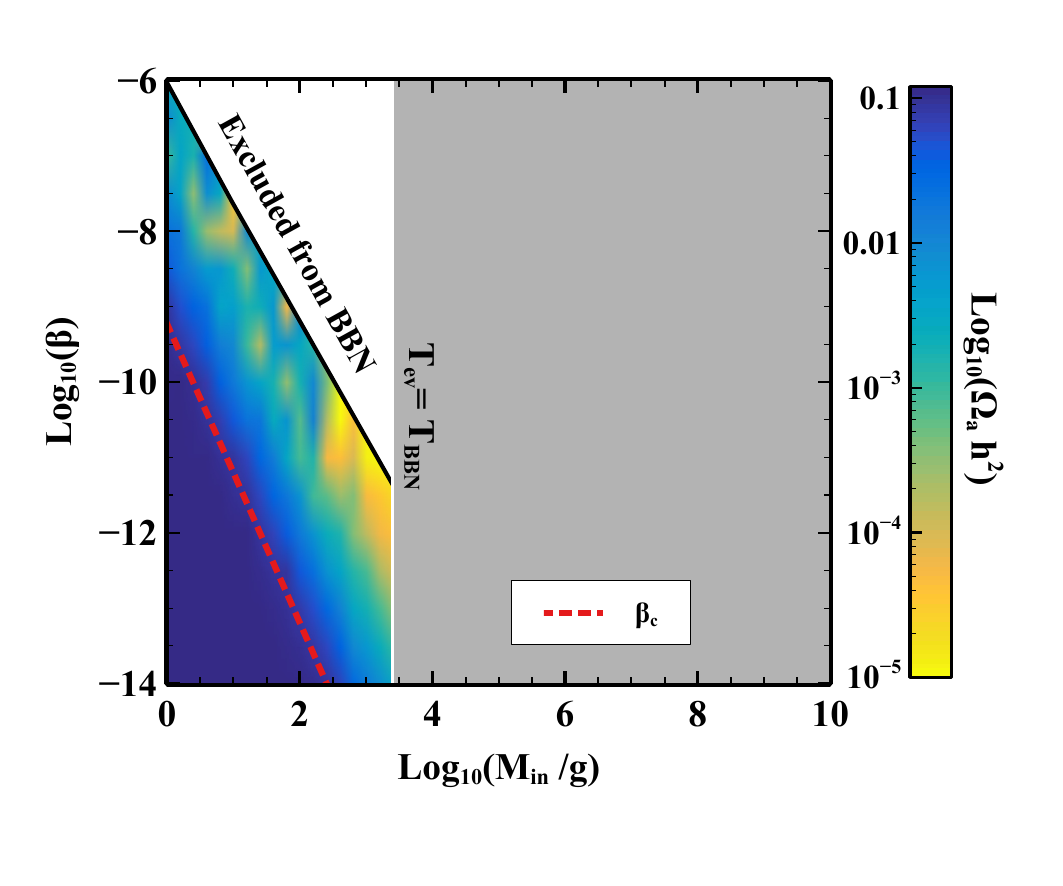}
    \caption{Variation of axion relic abundance from kinetic misalignment as a function of PBH parameters considering SC (left panel) and MB regime with $k=1, q=0.5$ (right panel). The initial velocity at $T=f_a$ is $\dot{\theta_i} \, = \,3 \times 10^{38} \, {\rm s}^{-1}$ whereas $f_a=10^{10}$ GeV.}
    \label{fig7}
\end{figure}

Fig. \ref{fig6} and Fig. \ref{fig7} show the variation of axion DM relic density for vacuum and kinetic misalignment scenarios respectively in the plane of PBH parameters keeping axion parameters fixed. The left and right panels of each of these figures correspond to SC and MB regimes respectively. While axion DM remains underproduced for the chosen axion parameters in vacuum misalignment case, correct relic can be produced in the kinetic misalignment cases for specific choices of PBH parameters. Both vacuum and kinetic misalignment mechanisms show decrease in axion DM relic in the presence of PBH for fixed axion parameters, as evident from these two figures.


\begin{figure}
\centering
\includegraphics[width=8cm]{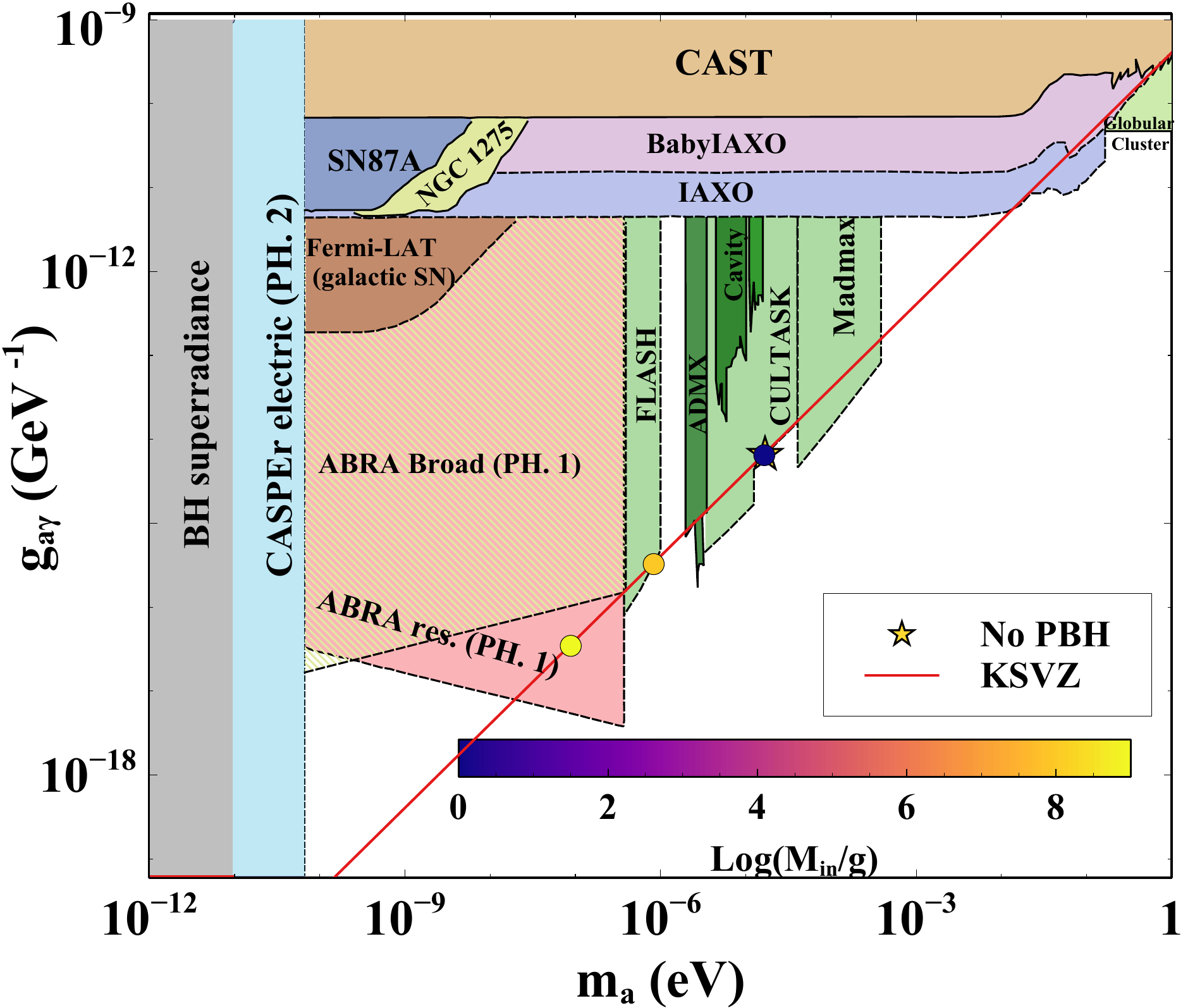}
\includegraphics[width=8cm]{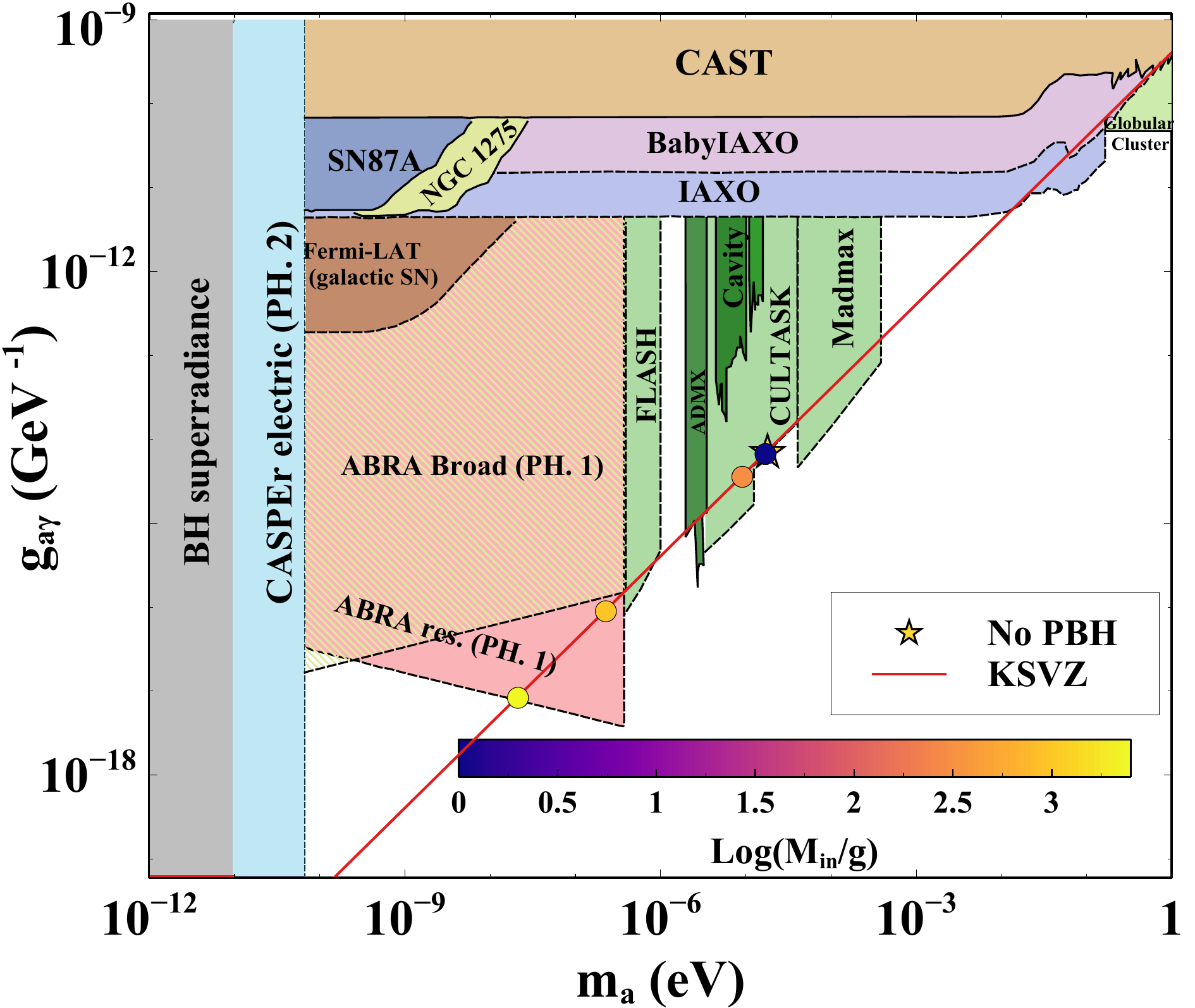}
    \caption{Axion-photon coupling versus axion mass in the presence of PBH with $\beta = 10^{-10}$ and varying $M_{\rm in}$ considering SC regime (left panel) and MB regime with $k=1, q=0.5$ (right panel). Axion DM relic is assumed to be generated via the vacuum misalignment mechanism with $\theta_i=\pi/\sqrt{3}$.}
    \label{fig8}
\end{figure}

\begin{figure}
\centering
\includegraphics[width=8cm]{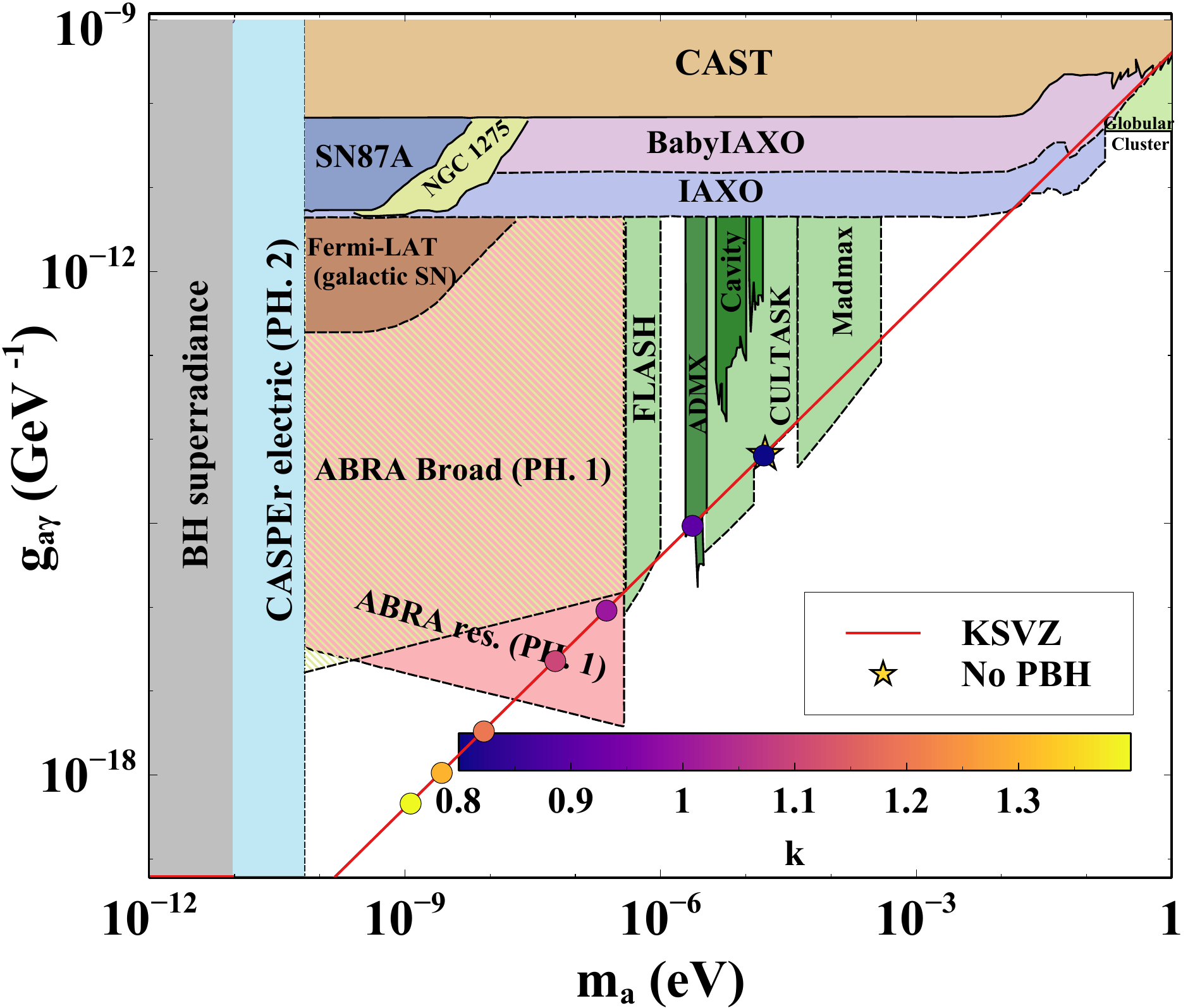}
\includegraphics[width=8cm]{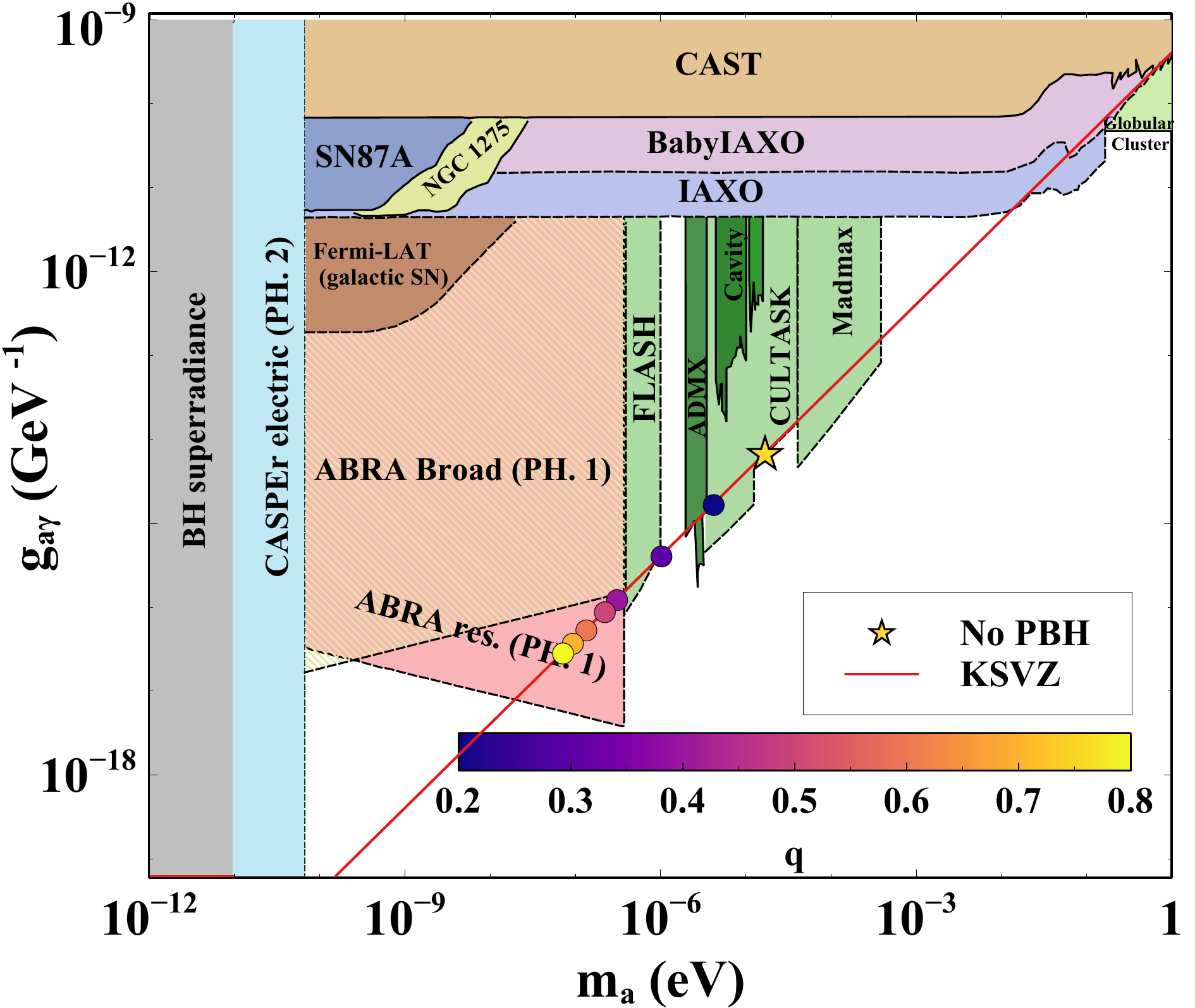}
    \caption{Axion-photon coupling versus axion mass in the presence of PBH with $M_{\rm in} = 10^3$ g and $\beta = 10^{-10}$ showing the impact of memory-burden on DM relic satisfying points in terms of varying $(k, q)$ with $q=0.5$ (left panel) and $k=1$ (right panel). Axion DM relic is assumed to be generated via the vacuum misalignment mechanism with $\theta_i=\pi/\sqrt{3}$.}
    \label{fig9}
\end{figure}
\begin{figure}
\centering
\includegraphics[width=8cm]{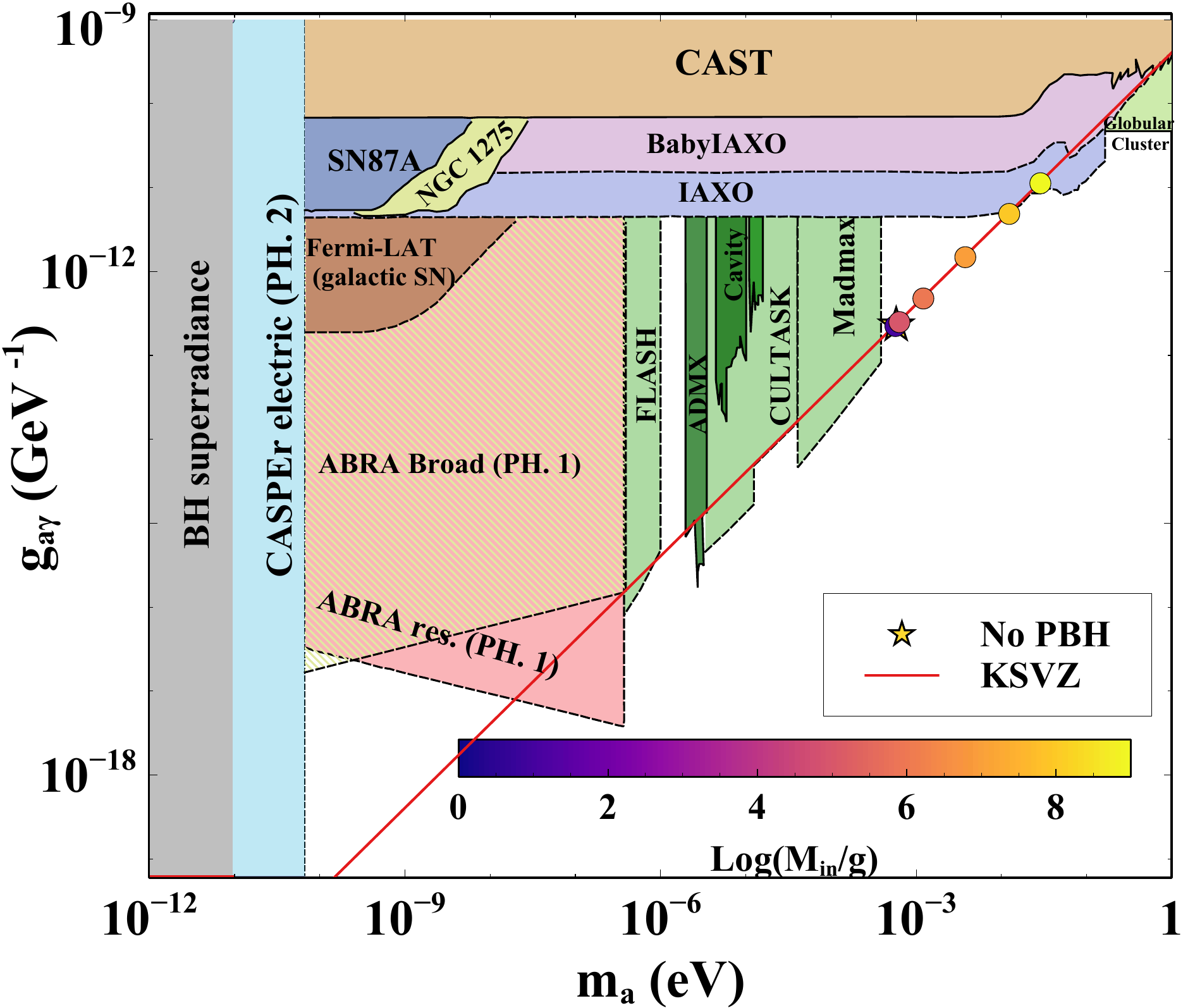}
\includegraphics[width=8cm]{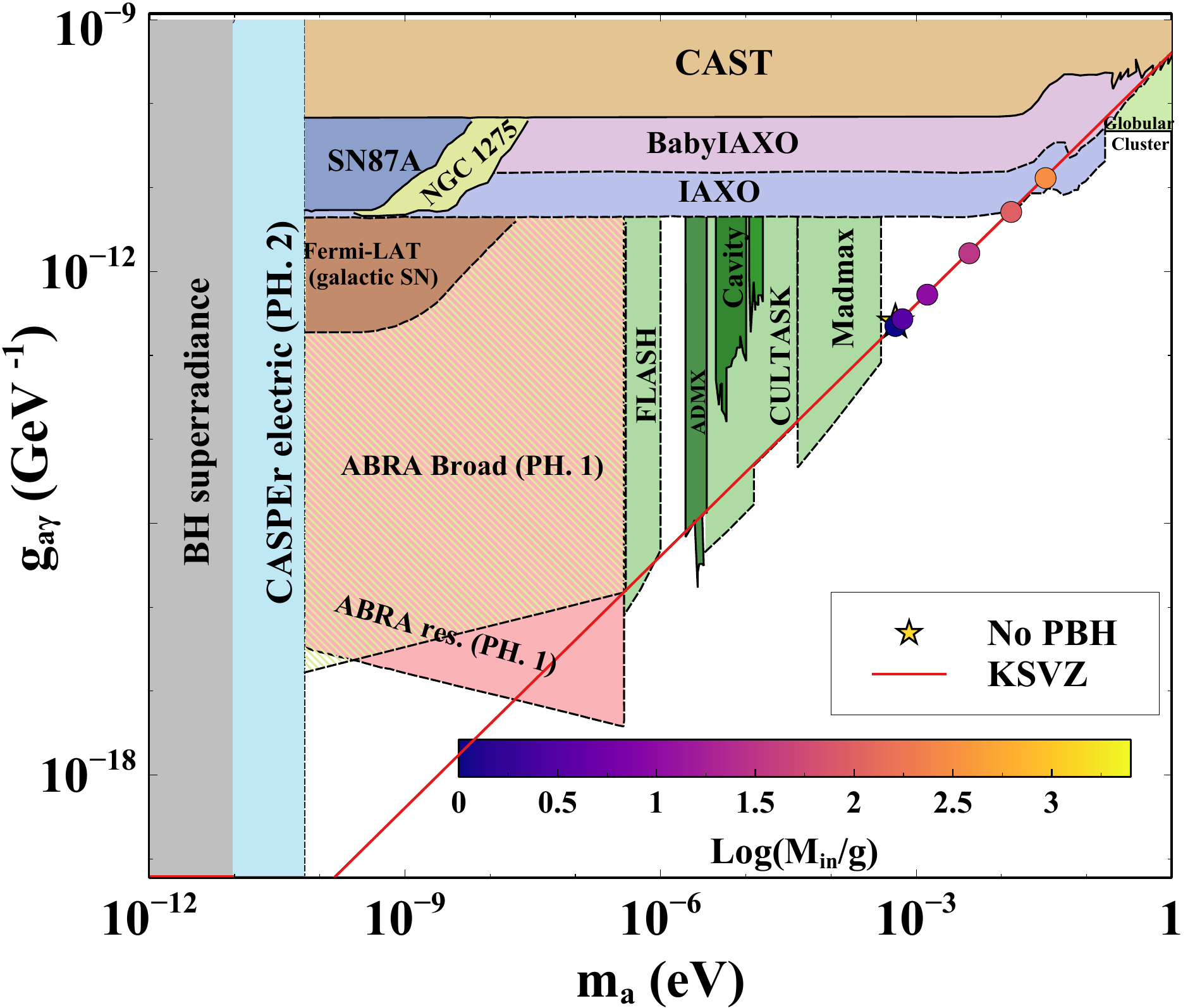}
    \caption{Axion-photon coupling versus axion mass in the presence of PBH with fixed $\beta = 10^{-10}$ and varying $M_{\rm in}$ considering SC regime (left panel) and MB regime with $k=1, q=0.5$ (right panel). Axion DM relic is assumed to be generated via the kinetic misalignment mechanism with $\dot{\theta}_i = \, 3 \times \,10^{38}\, {\rm s}^{-1}$ at $T=f_a$.}
    \label{fig10}
\end{figure}

\begin{figure}
\centering
\includegraphics[width=8cm]{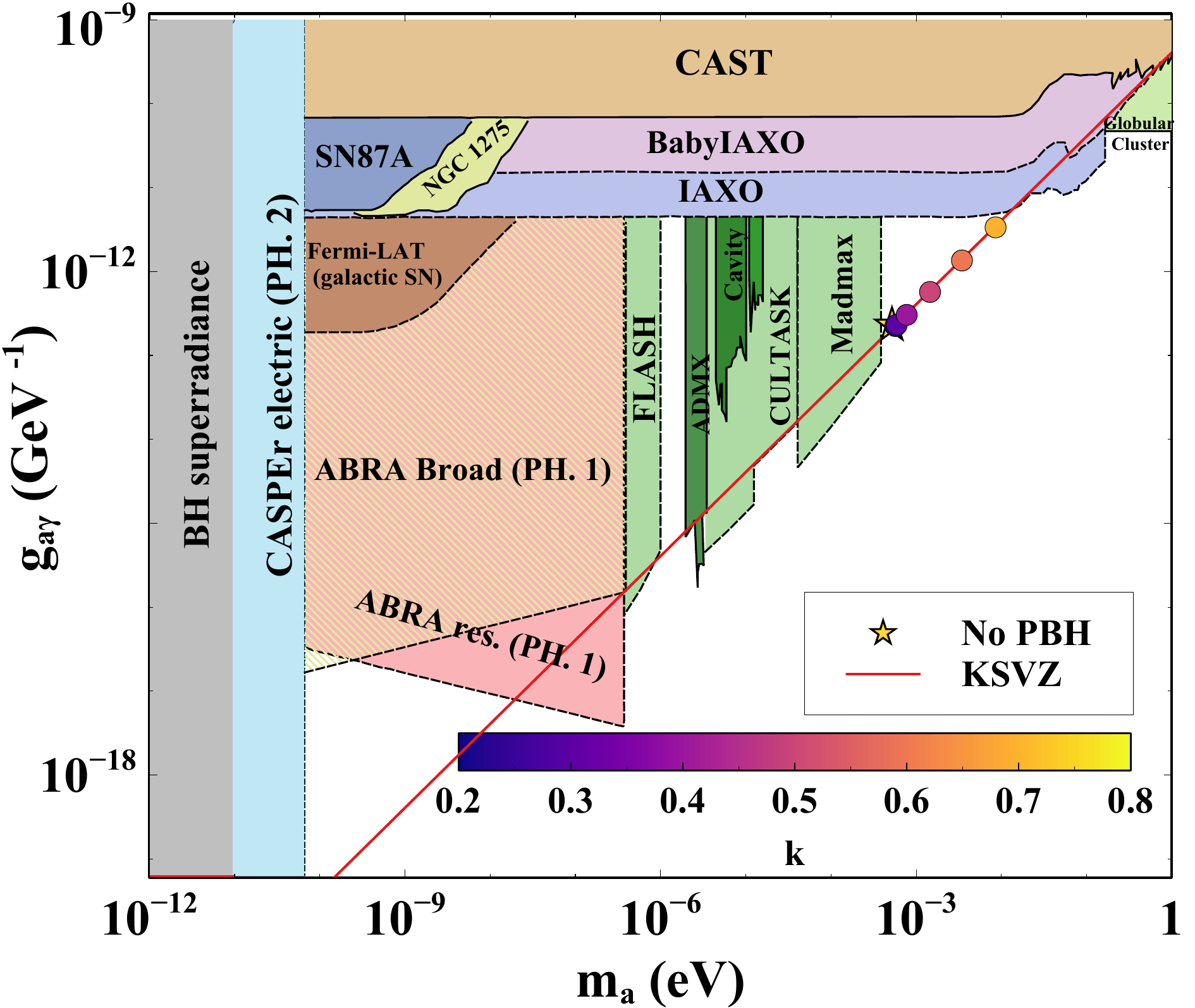}
\includegraphics[width=8cm]{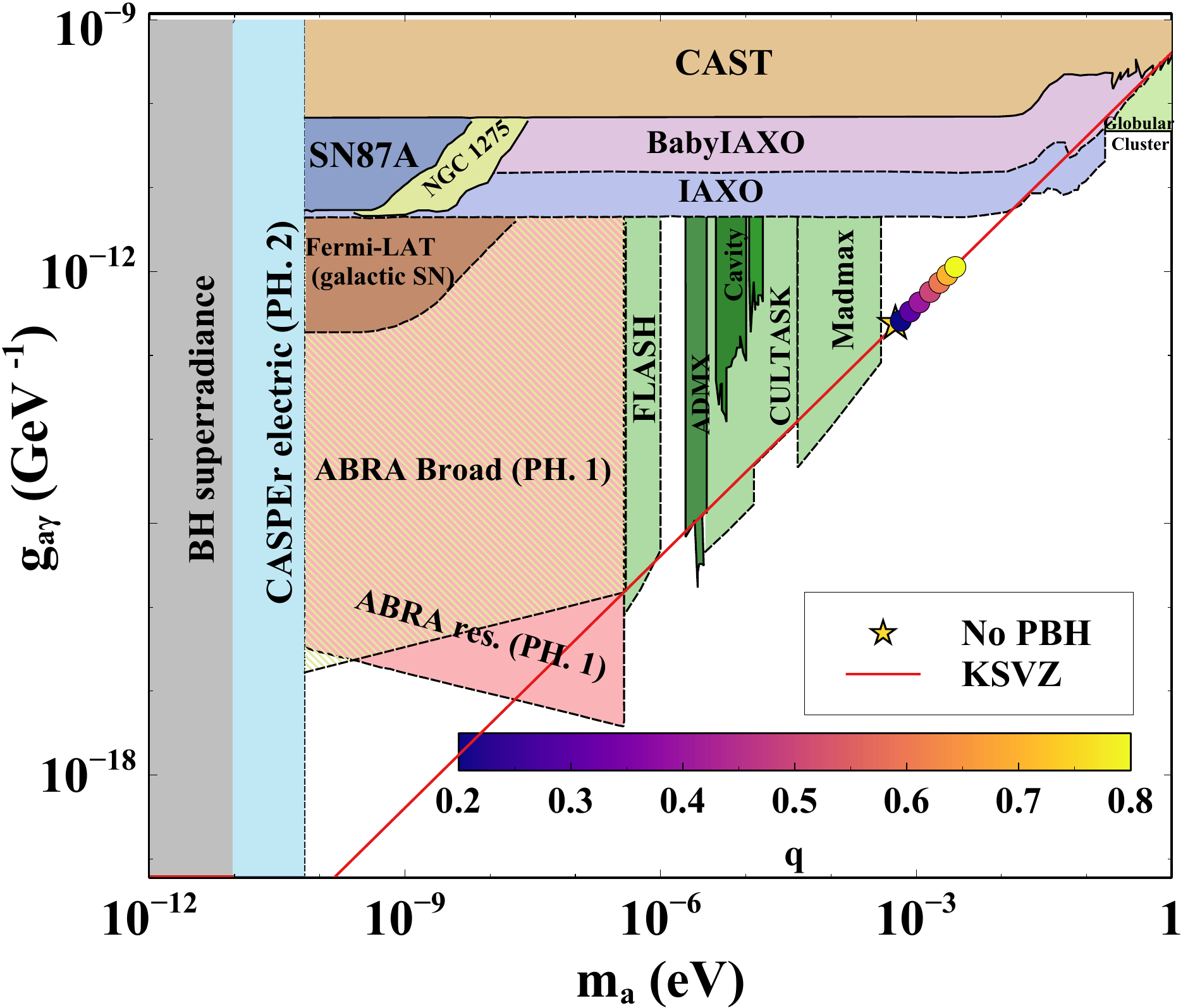}
    \caption{Axion-photon coupling versus axion mass in the presence of PBH with $M_{\rm in} =  10^3$ g and $\beta = 10^{-10}$ showing the impact of memory-burden on DM relic satisfying points in terms of varying $(k, q)$ with $q=0.5$ (left panel) and $k=1$ (right panel). Axion DM relic is assumed to be generated via the kinetic misalignment mechanism with $\dot{\theta}_i = \, 3 \times \,10^{38}\, {\rm s}^{-1}$ at $T=f_a$.}
    \label{fig11}
\end{figure}

\begin{figure}
    \centering
    \includegraphics[width=0.49\linewidth]{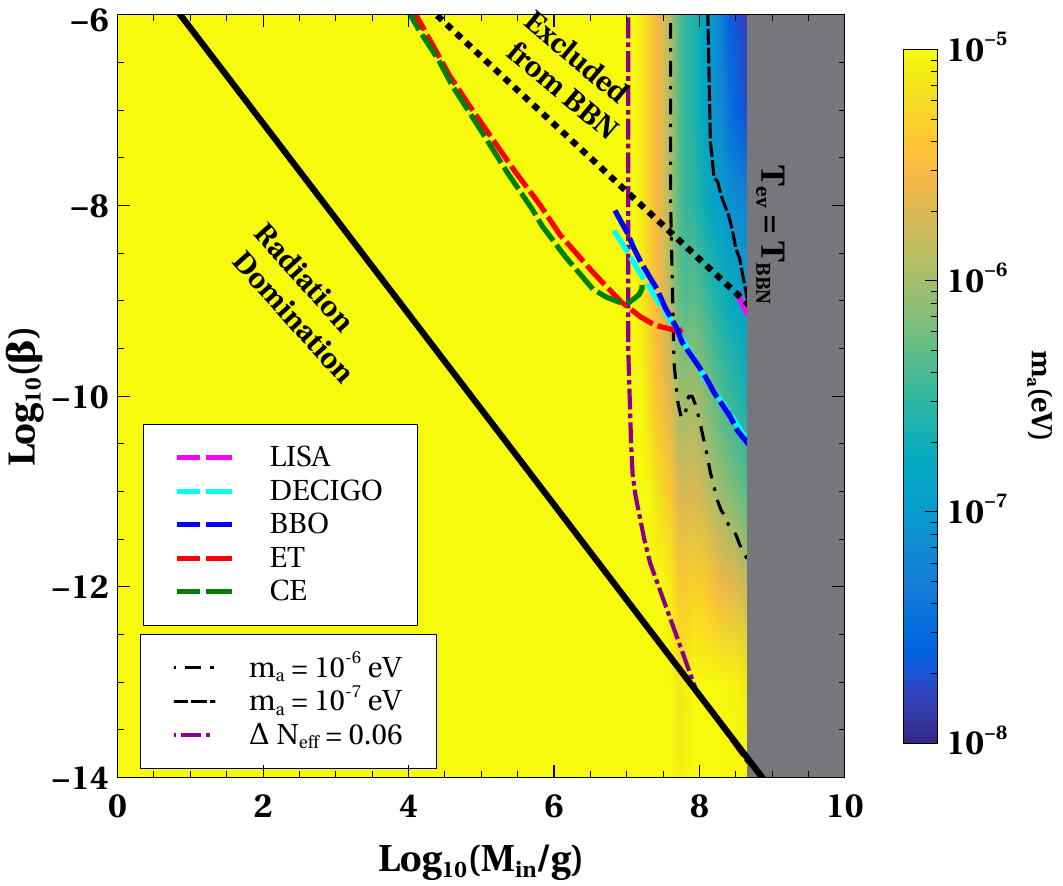}
    \includegraphics[width=0.49\linewidth]{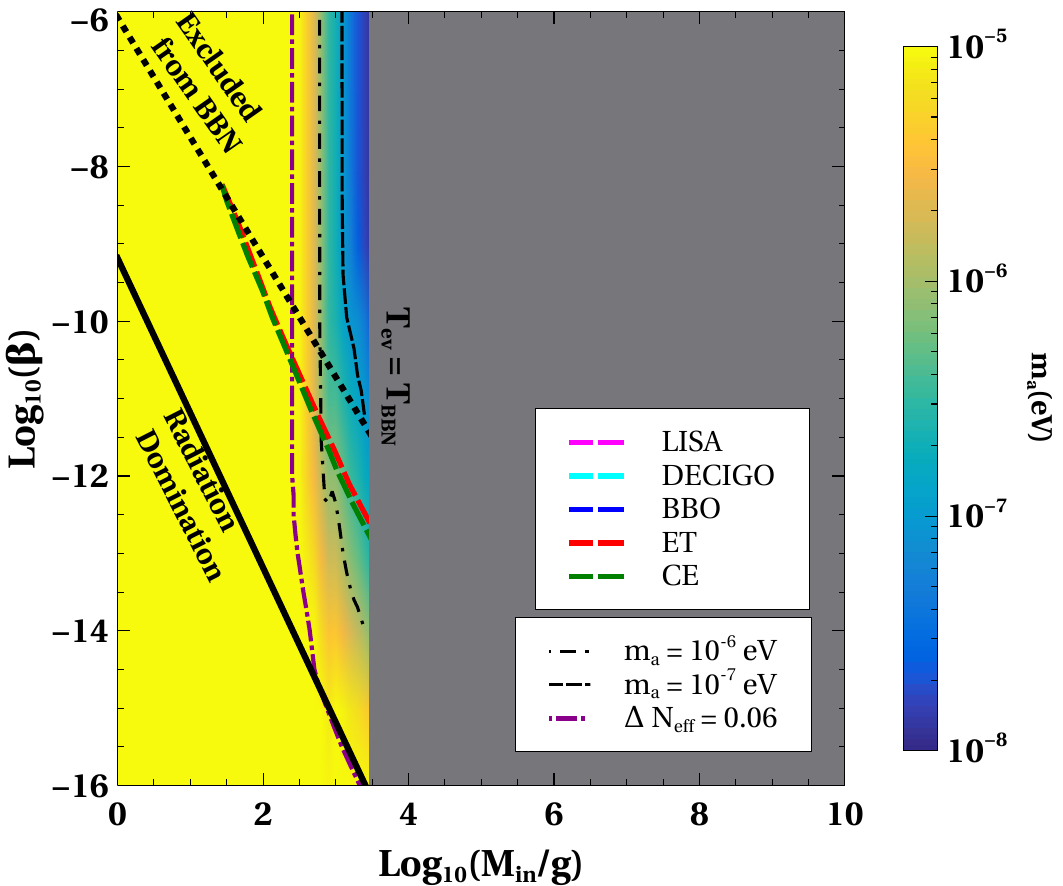}
    \caption{Variation of PBH parameter space $\beta$-$M_{\rm in}$ with axion mass in color bar satisfying DM relic density for SC regime (left panel) and MB regime with \{$k=1$, $q=0.5$\} (right panel). Dashed (dot-dashed) colored contours indicate sensitivities of future GW (CMB) experiments. Axion DM relic is assumed to be generated via the vacuum misalignment mechanism with $\theta_i=\pi/\sqrt{3}$.}
    \label{fig:summary_vac}
\end{figure}
\begin{figure}
    \centering
    \includegraphics[width=0.49\linewidth]{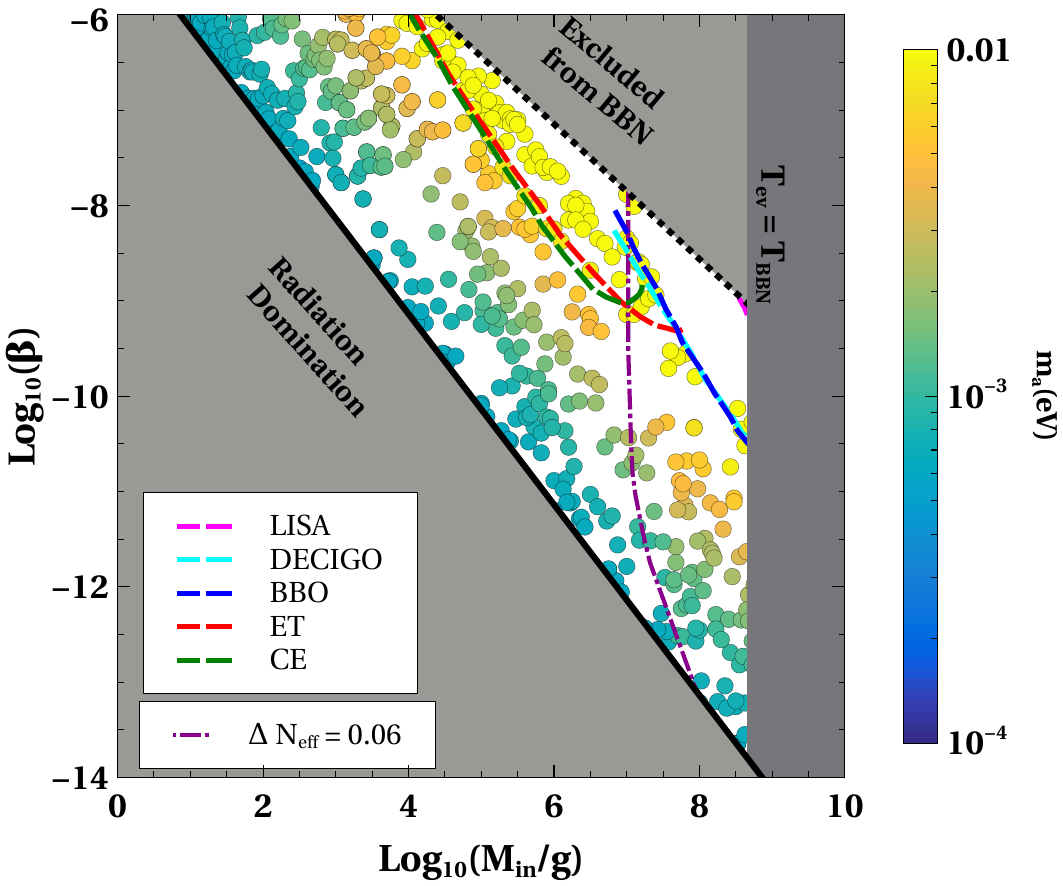}
    \includegraphics[width=0.49\linewidth]{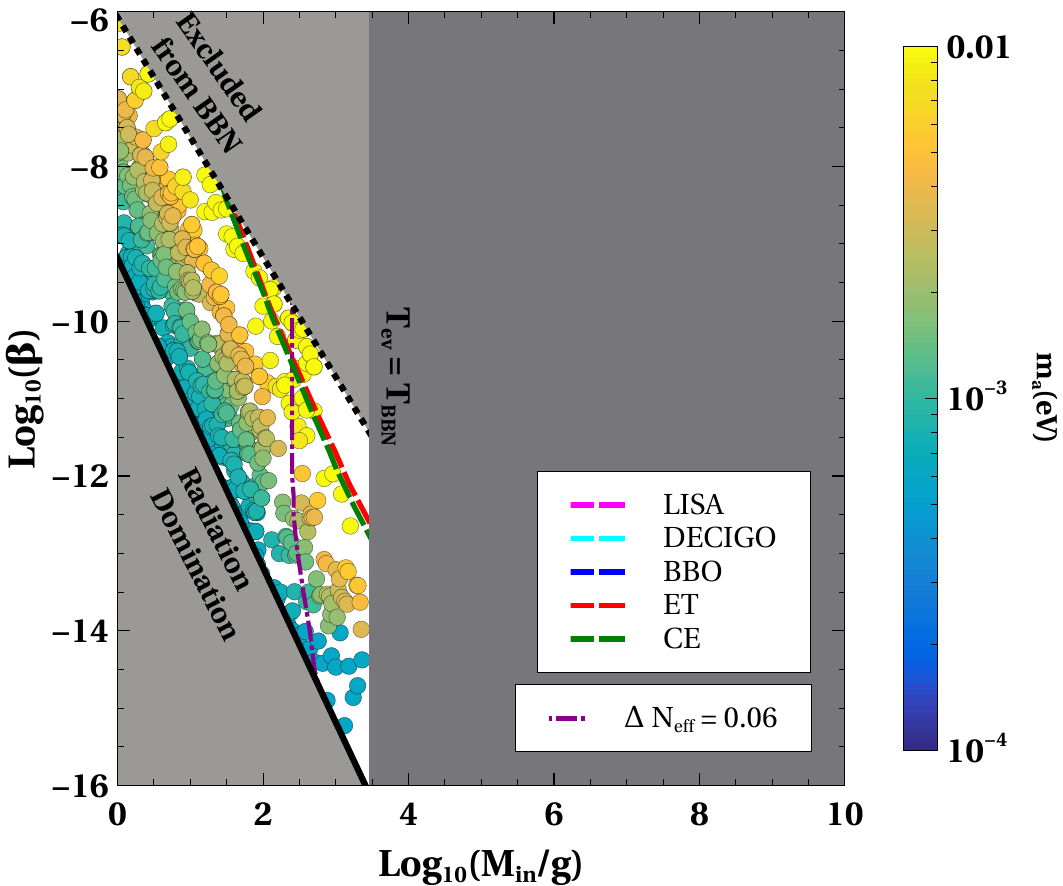}
    \caption{Variation of PBH parameter space $\beta$-$M_{\rm in}$ with axion mass in color bar satisfying DM relic density for SC regime (left panel) and MB regime with \{$k=1$, $q=0.5$\} (right panel). Dashed (dot-dashed) colored contours indicate sensitivities of future GW (CMB) experiments. Axion DM relic is assumed to be generated via the kinetic misalignment mechanism with $\dot{\theta}_i= 3 \times 10^{38}\, {\rm s}^{-1}$ at T = $f_a$.}
    \label{fig:summary_kin}
\end{figure}

\section{Detection aspects}
\label{sec5}
In this section we discuss the detection prospects of our scenario in typical axion detection experiments as well as gravitational wave and future CMB experiments.

\subsection{Axion detection}
Most of the axion detection experiments rely on the axion-photon coupling of the form $\frac{1}{4} g_{a\gamma} a F\Tilde{F}$ which for a KSVZ-type model can be found as \cite{Srednicki:1985xd}
\begin{eqnarray}
     g_{a\gamma} = -\frac{\alpha}{2 \pi f_{a}} \left(\frac{2}{3} \frac{4 m_{d}+m_{u}}{m_{u} + m_{d}}\right) = -1.92 \frac{\alpha}{2 \pi f_{a}},
 \end{eqnarray}
with $\alpha$ being the fine-structure constant. Fig. \ref{fig8}, \ref{fig9}, \ref{fig10} and \ref{fig11} show the parameter space in the plane of axion-photon coupling $g_{a\gamma}$ and axion mass $m_a$ including the bounds and sensitivities of different experiments or observables. The current experimental bounds on the axion-photon coupling from various experiments or observables are shown by the solid color lines (CAST \cite{CAST:2007jps, CAST:2017uph}, SN87A \cite{PhysRevLett.60.1797, PhysRevD.39.1020, PhysRevLett.60.1793}, NGC 1275 \cite{Fermi-LAT:2016nkz}, ADMX \cite{ADMX:2006kgb, Stern:2016bbw, ADMX:2019uok}, Globular clusters \cite{Ayala:2014pea}) whereas future experimental sensitivities or observables are shown by the dashed lines (CASPEr \cite{Budker:2013hfa}, FLASH \cite{Alesini:2017ifp,Alesini:2019nzq, Alesini:2023qed}, ABRACADABRA \cite{Kahn:2016aff, Ouellet:2018beu}, CULTASK \cite{Lee:2020cfj, Semertzidis:2019gkj}, MADMAX \cite{Caldwell:2016dcw}, IAXO \cite{Vogel:2013bta, IAXO:2019mpb},  Fermi-LAT \cite{Meyer:2016wrm}, BH superradiance \cite{Cardoso:2018tly}). Fig. \ref{fig8} and \ref{fig9} show the $g_{a\gamma}-m_a$ parameter space for vacuum misalignment in the presence PBH pointing out the differences in SC and MB regimes. One can clearly see the shift in parameter space towards lighter axion mass window due to the presence of PBH compared to the standard relic satisfying point. In case of vacuum misalignment, as we increase PBH Mass for fixed $\beta$, the oscillation starts earlier as seen from the top left panel of Fig. \ref{fig3}. This and the entropy injection from PBH evaporation leads to further decrease in axion abundance. This can be compensated by increasing $f_a$ and hence with a lighter axion mass. Fig. \ref{fig8} shows the decrease in axion mass satisfying correct relic as PBH mass increases. Fig. \ref{fig9} shows the corresponding plots in MB regime while keeping $M_{\rm in}, \beta$ fixed. As the values of $(k, q)$ increase, the oscillation begins at a lower temperature which was shown in the bottom panel of Fig. \ref{fig3}. While this could increase axion abundance for given PQ scale and PBH parameters, the entropy dilution due to PBH evaporation can lower the relic as well. As it turns out, the entropy dilution effect dominates over the effect of changing oscillation temperature, leading to decrease in axion abundance with increase in $(k, q)$. Therefore, in order to get the desired axion abundance for a choice of PBH parameters, $f_a$ has to be increased which corresponds to a lower axion mass.

On the contrary, for kinetic misalignment scenario, the presence of PBH shift the relic allowed parameter space towards the heavier axion mass window compared to the relic satisfying point in a radiation dominated Universe, as seen from Fig. \ref{fig10} and \ref{fig11}. Here, as we increase PBH mass for a given $\beta$, the oscillation begins at a higher temperature as seen in the top left panel of Fig. \ref{fig5}. This, together with entropy dilution from PBH evaporation lead to reduced axion abundance. Since a decrease in PQ scale $f_a$ delays the oscillation in kinetic misalignment scenario, we can decrease $f_a$ thereby increasing axion mass to compensate for the effects due to PBH. This is clearly seen from Fig. \ref{fig10}. Now, for fixed ($M_{\rm in}, \beta$), if we increase the MB parameters $(k, q)$, oscillation begins at a higher temperature as seen from the bottom panel of Fig. \ref{fig5}. Once again, this leads to a decrease in axion abundance which gets lowered further due to entropy dilution. This can be compensated by considering a lower $f_a$ and hence heavier axion mass, as shown in Fig. \ref{fig11}.

 As we have discussed above, the vacuum misalignment parameter space shifts towards lower axion mass while the kinetic misalignment parameter space moves to the higher mass window in the presence of PBH. Therefore, the two different misalignment mechanisms offer interesting axion detection complementarities in the presence of PBH considering both SC and MB regimes of PBH evaporation.

\subsection{Gravitational waves}   
Presence of PBH in the early Universe can be associated with the production of gravitational waves in a variety of ways. For ultra-light PBH of our interest, we consider the production of GW from the density perturbations due to inhomogeneous distribution of PBH \cite{Papanikolaou:2020qtd, Domenech:2020ssp, Inomata:2020lmk}. This is not only independent of the details of PBH formation mechanism, but also leads to observable GW spectra having peak frequencies within reach of present and near future detectors. The inhomogeneous spatial distribution of PBH generates isocurvature density fluctuations. After PBH domination, these isocurvature perturbations are converted into adiabatic perturbations which induce GW at second order. These GW are further enhanced during the PBH evaporation \cite{Papanikolaou:2020qtd, Domenech:2020ssp}. 

The peak amplitude of GW from memory-burdened PBH at the epoch of evaporation can be expressed as \cite{Balaji:2024hpu, Barman:2024iht}
\begin{eqnarray}
    \Omega^{\rm peak}_{\rm {GW, ev}} &\simeq& \frac{1}{4133^{\frac{4}{3+2k}}}q^4\left(\frac{3+2k}{3}\right)^{-\frac{7}{3}+\frac{4}{9+6k}} \frac{\beta^{16/3}\,{\rm exp[8k(7-\frac{4}{3+2k})]}}{2.3\times10^{-20}}  \nonumber \\
     &\times& \left(\frac{q M_{\rm in}}{1 \rm g}\right)^{\frac{2}{3}(1+k)(7-\frac{4}{3+2k})} 
     \begin{cases}
          1, \text{ \hspace{0.5cm} for \hspace{0.5cm}} \beta >  \beta_{*} \hspace{0.5cm} \\ 
          q^8, \text{\hspace{0.5cm} for \hspace{0.5cm}} \beta <  \beta_{*}.
     \end{cases}
\end{eqnarray}
The quantity $\beta_{*}$ represents that particular value of $\beta$ above which MB effect is activated during PBH dominated era. This is given by
\begin{eqnarray}
    \beta_{*} = \left(\frac{3 \, \epsilon}{16 \, \pi \, \gamma (1-q^3) S(M_{\rm in})}\right)^{1/2} \simeq 7.3\times10^{-6} \frac{1}{\sqrt{1-q^3}} \left(\frac{1 \, \rm g}{M_{\rm in}}\right).
\end{eqnarray}
Incorporating the redshift of GW amplitude, the full spectrum today can be written as
\begin{eqnarray}
    \Omega_{\rm GW,0} h^2 (f) = 1.62\times 10^{-5} \, \Omega_{\rm GW, ev}^{\rm peak} h^2 \left(\frac{f}{f_{\rm UV}}\right)^{\frac{11+10k}{3+2k}} \mathcal{I}(f,k),
\end{eqnarray}
where 
\begin{eqnarray}
    \mathcal{I}(f,k) = \int_{-\xi_{0}(f)}^{\xi_{0}(f)} ds \frac{(1-s^2)^{2}}{ (1-c^2_{s} s^{2})^{(1+\frac{2}{3+2k})}},
\end{eqnarray}
and the quantity $\xi_{0}(f)$ can be read as
\begin{eqnarray}
    \xi_{0}(f) = 
    \begin{cases}
        1, \text{\hspace{0.5 cm} for \hspace{0.5 cm}} \frac{f_{\rm UV}}{f} \geq \frac{1+c_{s}}{2 c_{s}} \\
        \frac{2 f_{\rm UV}}{f} - \frac{1}{c_{s}}, \text{\hspace{0.5 cm} for \hspace{0.5 cm}} \frac{1+c_{s}}{2 c_{s}} \geq \frac{f_{\rm UV}}{f} \geq \frac{1}{2 c_{s}} \\
        0, \text{\hspace{0.5 cm} for \hspace{0.5 cm}} \frac{1}{2 c_{s}} \geq \frac{f_{\rm UV}}{f}.
    \end{cases}
\end{eqnarray}
Here, $c_{s}$ denotes the sound speed and takes value of $\frac{1}{\sqrt{3}}$ during radiation dominated era and the frequency related to the cutoff scale is
\begin{eqnarray}
    f_{\rm UV} \simeq 4.8\times 10^{6}\, \rm{Hz} \, e^{-4k} \left(\frac{3+2k}{3}\right)^{1/6} \left(\frac{1\, \rm g}{q \, M_{\rm in}}\right)^{\frac{5}{6}+\frac{k}{3}}.
\end{eqnarray}
For $\beta > \beta_{*}$, we only consider $q$ values larger than $0.41$. The reason being, for $q<0.41$ with $\beta > \beta_{*}$, another intermediate radiation domination arise which is not considered for the derivations of GW spectrum \cite{ Balaji:2024hpu,Barman:2024iht}.

Fig. \ref{fig:summary_vac} and \ref{fig:summary_kin} show summary of the allowed parameter space in PBH parameter space $\beta-M_{\rm in}$ with axion mass in color bar for vacuum and kinetic misalignment scenarios respectively. The GW sensitivities of future GW detectors BBO~\cite{Crowder:2005nr,Corbin:2005ny,Harry:2006fi}, DECIGO~\cite{Seto:2001qf,Kawamura:2006up,Yagi:2011wg}, CE~\cite{LIGOScientific:2016wof,Reitze:2019iox}, ET~\cite{Punturo:2010zz, Hild:2010id,Sathyaprakash:2012jk, Maggiore:2019uih}, LISA~\cite{2017arXiv170200786A} are shown as dashed colored contours. We derive the sensitivity of future GW detectors to the entire parameter space by focusing on the peak amplitude ($\Omega^{\rm peak}_{\rm GW, ev}$) and cut-off frequency ($f_{\rm UV}$) instead of the full GW spectrum. It should be noted that the peak amplitude is a function of PBH parameters $\Omega^{\rm peak}_{\rm GW}=\Omega^{\rm peak}_{\rm GW}(M_{\rm in}, k, q, \beta$) and the cut-off frequency $f_{\rm UV}=f_{\rm UV}(M_{\rm in}, k, q)$ does not depend upon $\beta$. Fixing the values of $k$ and $q$, it is possible to project the GW sensitivities into the $M_{\rm in}$-$\beta$ plane shown in Fig. \ref{fig:summary_vac} and \ref{fig:summary_kin}. While the left-panels of these figures show the SC regime, the right panels consider the MB regime with $k=1, q=0.5$. Evidently, the correlations among PBH parameters and axion mass change significantly depending upon the type of misalignment mechanism and evaporation regimes of the PBH. 

\subsection{CMB signatures} 
Apart from the production from misalignment mechanism, axions are also produced from PBH evaporation. In principle, the energy density of axion produced from PBH evaporation depends on PBH initial mass $M_{\rm in}$ and initial fractional energy density $\beta$ along with axion mass $m_{a}$. As axion mass is negligibly small compared to the instantaneous Hawking temperature of black hole in our scenario, the produced axion energy density from PBH is nearly independent of axion mass. While axions produced from misalignment mechanism act as cold DM, the axions produced from PBH evaporation act as hot DM or dark radiation and contribute to the Hubble expansion rate as radiation. Future CMB experiments can measure such extra radiation energy density of the early Universe to a very high accuracy and can constrain a portion of PBH parameter space. 

The contribution of dark radiation to effective number of relativistic species can be parameterised as 
\begin{eqnarray}
    N_{\rm eff} = \frac{\rho_{\rm DR}(T_{\rm eq})}{\rho_{\rm R} (T_{\rm eq})} \left[N^{\rm SM}_{\rm eff} + \frac{8}{7}\left(\frac{4}{11}\right)^{-4/3}\right],
\end{eqnarray}
where $N^{\rm SM}_{\rm eff}$ takes value of $3.044$ \cite{Akita:2020szl, Bennett:2020zkv, Froustey:2020mcq} in the SM and $T_{\rm eq}$ denotes the temperature at standard matter-radiation equality. In terms of $\Delta N_{\rm eff} = N_{\rm eff} - N^{\rm SM}_{\rm eff}$ and PBH evaporation temperature, the above equation can be written as 
\begin{eqnarray}
    {\rm \Delta N_{\rm eff}} = \left\{\frac{8}{7}\left(\frac{4}{11}\right)^{-\frac{4}{3}}+{\rm N_{\rm eff}^{\rm SM}}\right\} 
    \frac{\rho_{\rm DR}(T_{\rm ev})}{\rho_{\rm R}(T_{\rm ev})}
    \left(\frac{g_*(T_{\rm ev})}{g_*(T_{\rm eq})}\right)
    \left(\frac{g_{*s}(T_{\rm eq})}{g_{*s}(T_{\rm ev})}\right)^{\frac{4}{3}}.
\end{eqnarray}
Here we numerically calculate $\rho_{\rm DR}$ at the evaporation temperature by solving the  following Boltzmann equation (together with Eq. \eqref{eqn:beq}, \eqref{eqn:beq1}, \eqref{eqn:beq2}, \eqref{eqn:beq3} and \eqref{eq:massloss2})
\begin{eqnarray}
    \frac{d \rho_{\rm DR}}{dt} + 4 \mathcal{H} \rho_{\rm DR} = - \frac{g_{\rm DR, H}(T_{\rm BH})}{g_{*,\rm H}(T_{\rm BH}) + g_{\rm DR, H}(T_{\rm BH})}\frac{1}{M_{\rm BH}} \frac{d M_{\rm BH}}{dt} \rho_{\rm BH},
\end{eqnarray}
where $g_{*,\rm H}\simeq 108$ and $g_{\rm DR,H} = 1.82$ for axion. 

The current bound on $\Delta N_{\rm eff}$ from Planck 2018 measurement reads $\Delta N_{\rm eff} < 0.285$ at $2\sigma$  \cite{Planck:2018vyg}. Future CMB experiments like CMB-S4 will be sensitive upto $\Delta N_{\rm eff} = 0.06$ \cite{Abazajian:2019eic} providing a new detection prospect of axion dark radiation from PBH evaporation. The sensitivity of CMB-S4 to the final allowed parameter space is shown as dot-dashed contour in Fig. \ref{fig:summary_vac} and \ref{fig:summary_kin}. Clearly, CMB-S4 can probe a part of the parameter space which can not be reached by the GW experiments discussed here, offering an interesting complementarity. 

\section{Conclusion}
\label{sec6}
We have studied the possibility of axion misalignment in a non-standard cosmological history where ultra-light primordial black holes dominate the early Universe. While vacuum misalignment mechanism in the presence of PBH was studied before, we have revisited it to include the memory-burden effects or backreaction of the emitted quanta on the black hole itself. Depending upon the parameters controlling the memory-burden effect, we show the shift in PBH as well as the QCD axion parameter space by comparing the results in standard cosmology, PBH with semi-classical evaporation and PBH with memory-burden effect. We then study the kinetic misalignment mechanism for the first time in the presence of PBH by considering both SC and MB regimes. As expected, the parameter space for both PBH as well as QCD axion change significantly in kinetic misalignment scenario. We then discuss the detection prospects of these four scenarios namely, vacuum and kinetic misalignment in SC or MB regimes of PBH in typical axion detection experiments as well as complementary probes offered by gravitational waves and CMB observations in future. We find a large portion of the sub-eV axion mass region consistent with dark matter relic that remains within reach of axion detection, future GW and future CMB experiments. 

\acknowledgements

The work of D.B. is supported by the Science and Engineering Research Board (SERB), Government of India grants MTR/2022/000575 and CRG/2022/000603. D.B. also acknowledges the support from the Fulbright-Nehru Academic and Professional Excellence Award 2024-25. The work of N.D. is supported by the Ministry of Education, Government of India via the Prime Minister's Research Fellowship (PMRF) December 2021 scheme.

\appendix

\section{PQ symmetry breaking during inflation}
\label{appen1}

So far, we have discussed the \textit{post-inflationary} scenario for PQ symmetry breaking. In the \textit{post-inflationary} scenario, the PQ symmetry is broken either after inflation or is broken during inflation, but it is restored again. Here, we discuss the \textit{pre-inflationary} scenario where PQ symmetry is spontaneously broken during inflation and is never restored afterwards. For PQ symmetry to be broken during inflation, axion decay constant must be larger than the Hubble expansion rate at the end of inflation, $f_{a} > \mathcal{H}_{\rm I}$. Additionally, if axion decay constant is larger than the maximum temperature reached in the post-inflationary era $f_{a} > T_{\rm max}$, the PQ symmetry is never restored again. When both of these conditions are satisfied, we get the \textit{pre-inflationary} scenario that gives a homogeneous value of the initial misalignment angle $\theta_{i}$. The \textit{pre-inflationary} scenario produces isocurvature perturbation that severely constrain the Hubble parameter during inflation, $\mathcal{H}_{\rm I}$.

Just like the inflaton field, the axion field present during inflation develops quantum fluctuations with typical standard deviation $\sigma_{a}$ given by
\begin{eqnarray}
    \sigma_{a} = \sqrt{\langle |\delta a(x)|^2 \rangle} \simeq \frac{\mathcal{H}_{\rm I}}{2 \pi}.
\end{eqnarray} 
The magnitude of isocurvature perturbations at a given length scale, $S_{\rm iso}$ is given as \cite{Beltran:2006sq, Bae:2008ue, Kobayashi:2013nva}
\begin{eqnarray}
    S_{\rm iso} = \frac{\Omega_{a}}{\Omega_{\rm DM}}\, \sigma_{a} \, \frac{\partial \, \text{ln} \,  \Omega_{a}}{\partial a}.
\end{eqnarray}
In our scenario, the axion constitutes the whole DM abundance $\Omega_{a} = \Omega_{\rm DM}$. Assuming $\Omega_{a} \propto \theta^2_{i}\, h(\theta_{i})$, where $h(\theta_{i})$ accounts for the anharmonicity in the axion potential  \cite{Strobl:1994wk,Visinelli:2009zm,Mazde:2022sdx}, the isocurvature power spectrum can be written as 
\begin{eqnarray}
    \Delta^2_{a}(k) = |S_{\rm iso}|^2 = \frac{\mathcal{H}^2_{\rm I}}{\pi^2\, f^2_{a}\, \theta^2_{i}} F(\theta_{i}), \hspace{1cm} \text{where}\,\,\, F(\theta_{i}) = \left(1+\frac{\theta_{i}}{2} \frac{\partial\, \text{ln}\, h(\theta_{i})}{\partial \theta_{i}}\right)^2.
\end{eqnarray}
Here we adopt the following expression for $h(\theta_{i})$,
\begin{eqnarray}
    h(\theta_{i}) = \left[\text{ln}\left(\frac{e}{1-\theta^2_{i}/\pi^2}\right)\right]^{7/6}.
\end{eqnarray}
As the axion fluctuations during inflation are independent of the quantum fluctuations of the inflaton field, the resulting isocurvature perturbations are uncorrelated with the adiabatic curvature perturbations. 

The relative amplitude of isocurvature perturbations at the pivot scale is constrained from CMB measurements and is given by
\begin{eqnarray}
    \beta_{\rm iso}(k_{0}) = \frac{\Delta^2_{a}(k_{0})}{\Delta^2_{a}(k_{0}) + \Delta^2_{\mathcal{R}}(k_{0})} < 0.035 \hspace{1 cm} \text{at}\,\,\,  k_{0} = 0.002 \,\, \text{Mpc}^{-1}.
\end{eqnarray}
The adiabatic power spectrum is fixed at $\Delta^2_{\mathcal{R}}(k_{0}) \simeq 2.1\times10^{-9}$ which gives the translated bound on $\mathcal{H}_{\rm I}$ as
\begin{eqnarray}
    \mathcal{H}_{\rm I} \lesssim 3\times 10^{-5} \, \frac{f_{a}\, \theta_{i}}{\sqrt{F(\theta_{i})}}.
\end{eqnarray}

\begin{figure}
    \centering
    \includegraphics[width=0.45\linewidth]{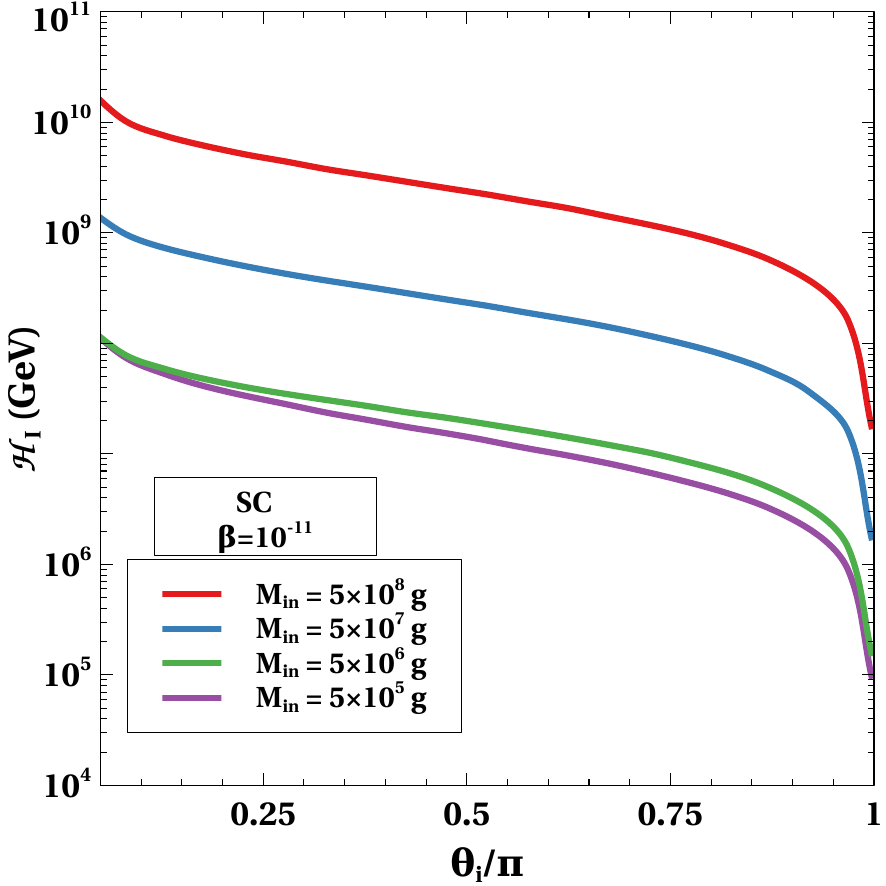}
    \includegraphics[width=0.45\linewidth]{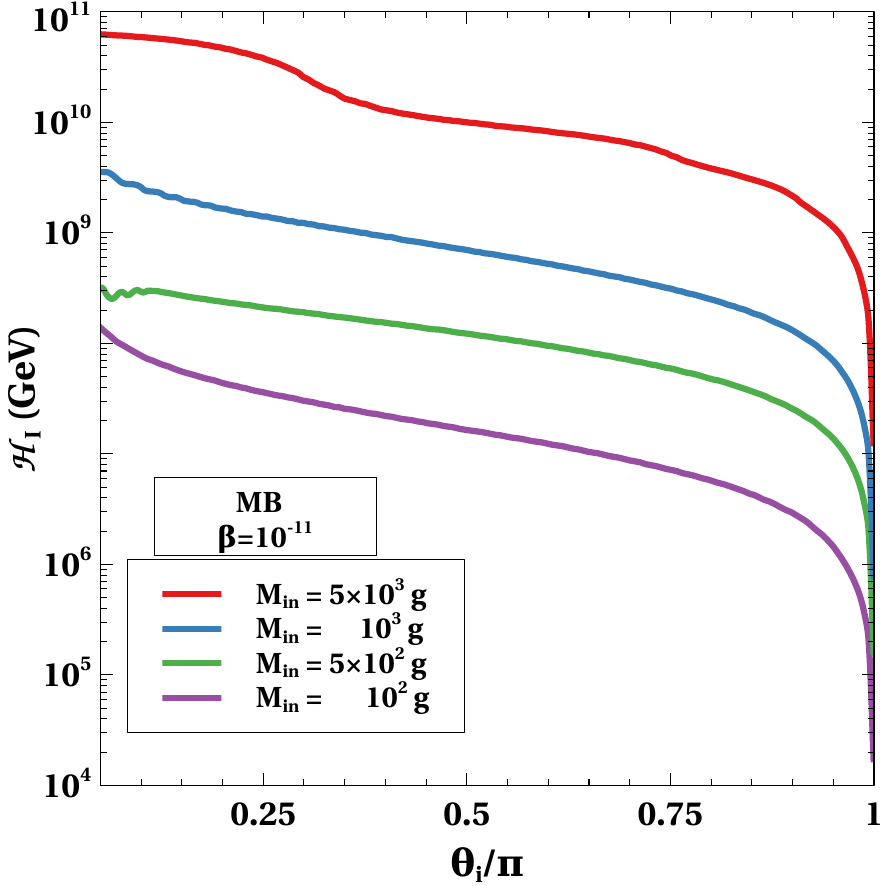}
    \caption{Constraint on $\mathcal{H}_{I}$ as a function of $\theta_{i}/\pi$ for different PBH masses assuming semi-classical regime (left) and memory-burdened regime (right). The regions above the solid line, for a given value of $M_{\rm in}$ is ruled out from CMB measurements. For both panels, $\beta$ is fixed at $10^{-11}$.}
    \label{fig:iso_cur}
\end{figure}

Fig. \ref{fig:iso_cur} shows the bound on $\mathcal{H}_{\rm I}$ as a function of $\theta_{i}$ for semi-classical (left panel) and memory-burdened (right panel) regimes. For MB regime, we fix $k=1, q=0.5$. For both the panels, $\beta$ is fixed at $10^{-11}$. Given a particular value of PBH initial mass $M_{\rm in}$, the regions above the solid line is ruled out from CMB measurements. The left panel indicates that presence of semi-classical PBH relaxes the bound on $\mathcal{H}_{\rm I}$ for PBH mass range $\sim$  \{$5\times 10^{5}$, $5\times10^{8}$\}g. For PBH masses below $\sim 5 \times 10^{5}$ g do not change the bound, as in this range PBH do not alter the evolution of the axion. The situation changes for PBH with memory-burden effect where a much lower PBH mass range $\sim $  \{$10^{2}$, $5\times10^{3}$\} 
g can relax the bound on $\mathcal{H}_{\rm I}$. As a result, memory-burdened PBH can alleviate the tension between high-scale inflation and axion isocurvature perturbations in different mass ranges.

While we have discussed the isocurvature constraints for the vacuum misalignment scenario in the presence of PBH in SC and MB regimes, similar bounds also exist for the kinetic misalignment scenario if PQ breaking occurs during inflation. Isocurvature constraints on kinetic misalignment scenario have been discussed in \cite{Co:2020jtv, Co:2022qpr} with possible remedies in \cite{Co:2023mhe}. Studying these constraints in the presence of PBH with or without memory-burden effect lie beyond the scope of this present work and is left for future studies.

%\bibliographystyle{JHEP}
%\bibliography{ref1} 

\providecommand{\href}[2]{#2}\begingroup\raggedright\endgroup

\end{document}